\documentclass[a4paper,11pt]{article}   
\usepackage{amsmath}   
\usepackage{graphicx}   
\usepackage{amssymb}   
\usepackage{latexsym,cite}   

\newlength{\epswid}   
\newcommand{\eq}{\begin{equation}}   
\newcommand{\en}{\end{equation}}   
\newcommand{\eqa}{\begin{eqnarray}}   
\newcommand{\ena}{\end{eqnarray}}

\newcommand{\dd}{{\mathrm{d}}}   
\newcommand{\ee}{{\mathrm{e}}}   
\newcommand{\ii}{{\mathrm{i}}}   
\newcommand{\bea}{\begin{eqnarray}}   
\newcommand{\eea}{\end{eqnarray}}

\begin{document}   
\begin{titlepage}   
\vskip0.5cm   
\begin{flushright}   
DFTT 29/05\\   
SISSA 87/2005/FM\\   
PTA/05-72\\   
MIT-CPT-3705\\   
gef-th-9/05\\   
\end{flushright}   
\vskip0.5cm   
\begin{center}   
{\Large\bf Potts correlators and the static three-quark potential}   
\end{center}   
\vskip0.5cm   
\centerline{   
M. Caselle$^{a}$, G. Delfino$^{b}$, P. Grinza$^{c}$, O. Jahn$^{d}$ and N.   
Magnoli$^{e}$}   
   
 \vskip0.4cm   
 \centerline{\sl  $^a$ Dipartimento di Fisica   
 Teorica dell'Universit\`a di Torino and I.N.F.N.,}   
 \centerline{\sl via P.Giuria 1, I-10125 Torino, Italy}   
 \centerline{\sl   
e--mail: \hskip 1cm   
 caselle@to.infn.it}   
 \vskip0.1 cm   
 \centerline{\em $^b$ International School for Advanced Studies (SISSA)}   
 \centerline{\em via Beirut 2-4, 34014 Trieste, Italy}   
 \centerline{\em INFN sezione di Trieste}   
 \centerline{\sl   
e--mail: \hskip 1cm delfino@sissa.it}   
\vskip0.1 cm   
 \centerline{\it $^{c}$ Laboratoire de Physique Th\'eorique et Astroparticules,   
Universit\'e Montpellier II,}   
 \centerline{ Place Eug\`ene Bataillon, 34095   
Montpellier Cedex 05, France}   
 \centerline{\sl   
e--mail: \hskip 1cm grinza@lpta.univ-montp2.fr}   
 \vskip0.1 cm   
 \centerline{\it $^{d}$Center for Theoretical Physics, MIT, Cambridge, MA 02139, USA}   
 \centerline{\sl   
e--mail: \hskip 1cm jahn@mit.edu}   
 \vskip0.1 cm   
 \centerline{\sl $^e$  Dipartimento di Fisica,   
 Universit\`a di Genova and   
 I.N.F.N.,}   
 \centerline{\sl via Dodecaneso 33, I-16146 Genova, Italy}   
 \centerline{\sl   
e--mail: \hskip 1cm magnoli@ge.infn.it}   
 \vskip0.4cm   
   
\begin{abstract}   
   
We discuss the two- and three-point correlators in the two-dimensional three-state Potts model  
in the high-temperature phase of the model. By using the form factor 
approach and perturbed conformal field theory methods we are able to describe both the large distance and the 
short distance behaviours of the correlators. We compare our predictions with a set of high precision Monte-Carlo 
simulations (performed on the triangular lattice realization of the model) 
finding a complete agreement in both regimes. In particular we use the two-point correlators to fix 
the various non-universal constants involved in the comparison (whose determination is one of the results of our 
analysis) and then use these constants to compare numerical results and theoretical predictions for the 
three-point correlator with no free parameter. Our results can be used to shed some light on the behaviour of the 
three-quark correlator in the confining phase of the (2+1)-dimensional 
SU(3) lattice gauge theory which is related by dimensional reduction to the three-spin correlator in the  
high-temperature phase of the three-state Potts model.  
The picture which emerges is that of a smooth crossover between a 
{\bf $\Delta$} type law at short distances and a {\bf Y} type law at large distances.

\end{abstract}   
\end{titlepage}   
   
\setcounter{footnote}{0}   
\def\thefootnote{\arabic{footnote}}

\section{Introduction}   
\label{sect1}   
The aim of this paper is to study the three-point correlator in the ${\mathbb{Z}}_3$   
Potts model outside the critical point. In particular we shall study the thermal   
perturbation of the model, in the high-temperature phase (i.e.\ the phase in   
which the symmetry is unbroken and the magnetization is zero).   
We shall first discuss the    
two-point function and compare it with high precision numerical simulations   
in order to fix all the normalizations and the   
non-universal quantities which appear in the three-point function. Then we shall   
address the three-point function and compare it again with numerical   
simulations. Thanks to the preliminary analysis of the two-point function this   
comparison will not require any fitting procedure but will be a direct and   
absolute comparison between theoretical predictions and numerical results.   
   
On the theoretical side we shall study  both the two- and the three-point   
correlators with two different tools.   
\begin{itemize}   
\item   
 The form factor approach which is   
essentially a large distance expansion. This approach requires  exact   
integrability, a condition which is indeed fulfilled by the scaling ${\mathbb{Z}}_3$ Potts model.   
\item   
The perturbative   
expansion around the conformal fixed point which is essentially a short   
distance expansion and is completely general, meaning that no specific   
integrability property of the perturbation under study is needed.    
\end{itemize}   
   
\noindent   
~From a theoretical point of view this is a rather interesting challenge:   
\begin{itemize}   
\item   
In the large distance regime (where the form factor approach is expected   
to hold) the strong coupling expansion suggests the existence of an additional    
 midpoint, inside the triangle spanned by the three spins of the correlators,   
 where the strong coupling paths emerging from the three spins converge and   
 join. The appearence of this new point is a novelty with respect to the   
 well known form factor calculation for the two point functions.    
 The way in which it is obtained is non-trivial and is one of the   
 interesting features of our analysis.   
  
\item   
Similarly it is the first time that the approach of   
the perturbative, infrared safe, expansion around the conformal solution   
is used for a three point function. This required some non-trivial extension of   
the techniques used in previous works on the two-point   
functions.   
   
\end{itemize}   
~From a physical point of view the scenario which emerges (which strongly   
resembles what one finds when looking at the three-quark potential in lattice   
gauge theories (LGT), a correspondence   
 which we shall discuss in detail below)  
 is a smooth crossover, as the distance among the three points increases,   
from a short distance behaviour in which the three point function is   
dominated by the three spin-spin interactions  along the edges of the triangle to   
a large distance behaviour in which the strong coupling expectation (the three   
spins joined by a path of minimal length) is fully realized.

This scenario is confirmed by the numerical simulations which turn out to be in   
remarkable agreement with our theoretical results.      
Indeed   
the important consequence of having exact analytic results    
for the two expansions is that we    
are able to compare our predictions with triangles of any size, both smaller and larger than the correlation length. Moreover,    
even if in the following we shall mainly perform our comparisons in the   
equilateral case, we can obtain analytic predictions for triangles of any   
shape,   
 thus allowing a   
very selective test of our results.    
 
Besides a better understanding of the three-state Potts model   
a second important motivation which we had in mind addressing the present problem  
is to use our results to shed some light on the behaviour of the three-quark potential in SU(3)  
LGT's. Indeed by dimensional reduction the (2+1)-dimensional SU(3) LGT can be mapped into the two-dimensional  
three-state Potts model and in particular  
the three-quark free energy is mapped into the three-point function of 
the Potts model (we shall discuss in detail this correspondence in sect.\ref{sect7} below). An important open 
problem in the study of the three-quark potential is to understand if the three-quark correlator follows 
the so called ``{\bf Y}'' or  ``{\bf $\Delta$}''  
law (for a discussion of these laws see again sect.\ref{sect7} below). Our 
results on the three-state Potts model suggest that the right picture is  
a smooth crossover between the two laws. 
At short distance, i.e.\ for interquark distances smaller than the correlation length, the three-quark potential 
is well described by the {\bf $\Delta$} law, which is indeed exact at the critical point (i.e.\ when the correlation 
length goes to infinity), while at large distances the correct description is the {\bf Y} law, which becomes 
exact in the strong coupling limit, i.e.\ for interquark distances much larger than the correlation length.  
This mixed behaviour agrees with the results of some recent simulations performed directly in the gauge 
model~\cite{fj05}.

This paper is organized as follows. Sect.\ref{sect2} contains a general   
discussion of the three-state Potts model both on the lattice and in the   
continuum and its CFT description at the critical point. In sect.\ref{sect3}    
we discuss the form factor analysis of the large distance expansion both for the   
two- and the three-point functions. Then in sect.\ref{sect4} we shall compare   
these predictions with a set of high precision Monte-Carlo simulations performed   
on a triangular lattice. Sect.\ref{sect5} is then devoted to the study of the   
short distance perturbative expansion which is then compared with the Monte-Carlo   
data in sect.\ref{sect6}. In sect.\ref{sect7} we shall then briefly comment on   
the implications of our results for the study of barionic states in LGT. 
Two appendices conclude the paper. The first contains some technical steps we omit in the main text, 
while the second is devoted to universal amplitude ratios.   
   
\section{The three-state Potts model}   
\label{sect2}   
The two-dimensional, isotropic, nearest-neighbor, three-state Potts model at temperature $T$    
on a lattice $\Lambda$ is defined by the partition function   
\eq   
Z = \sum_{ \{s_n \} } e^{-\beta {\cal H}}   
\label{eq1}   
\en   
with the Hamiltonian   
\eq   
{\cal H} = J \sum_{\langle nn' \rangle}(1-\delta_{s_n s_{n'}})   
\label{eq2}   
\en   
where $s_n=0,1,2$ are ${\mathbb Z}_3$-valued variables on each    
site\footnote{For this reason we shall often denote in the following the model as the   
 ${\mathbb{Z}}_3$ Potts model.},   
  $n \in \Lambda$, $\beta = (k_BT)^{-1}$, and $\langle n n' \rangle$   
denotes pairs of nearest-neighbor sites~\cite{potts}. The symmetry group of the Potts    
Hamiltonian is $S_3$, the permutation group of $3$ objects. In the following we   
shall study in particular the case in which $\Lambda$ is a   
regular triangular lattice with honeycomb boundary conditions.   
     
This choice is   
motivated by the fact that in the following we shall mainly be interested in 
the three-point function of the model, for  which a more symmetric choice of   
arguments is possible on a triangular lattice.   
 Notice however that most of our theoretical results are   
obtained in the continuum limit theory and hence, once the non-universal, lattice dependent,   
 normalizations are fixed, hold   
for any lattice $\Lambda$. A general introduction to the Potts model can be found in   
the review by Wu~\cite{wurev}. Recently a set of interesting results (series   
expansions and exact free energy calculation on strips) for the triangular   
lattice realization of the model in which we are interested appeared    
in a series of papers (see in particular~\cite{fgjsh},\cite{cjss} and references   
therein) to which we refer the interested reader.   
   
The model is known to have a second order phase transition for a critical value   
$T_c$ of the temperature which separates the high-temperature, symmetric phase   
from the low-temperature one in which the symmetry is broken and a spontaneous   
magnetization appears. In the following we shall be interested in the symmetric   
high-temperature phase. This is due to the fact that (as we shall discuss   
in the last section) we plan to use our results to better understand the   
behaviour of the baryonic states in LGT, and the confining phase of the SU(3) LGT   
is  mapped by dimensional reduction to the high-temperature phase of the   
$3$-state Potts model.   
   
Assuming that a single phase transition point exists in the model, the critical   
temperature can be obtained exactly even for the triangular lattice by   
using duality and the star triangle relation which, combined together, lead to an 
algebraic equation for $T_c$~\cite{beta_c}. Defining  $a = e^{\beta}$ we have   
\eq   
a^3-3a-1=0   
\en   
whose unique positive solution is    
\eq   
a = \cos(2\pi/9)+\sqrt{3}\sin(2\pi/9)=1.879385...   
\en   
to which the critical value $\beta_c=0.6309447...$ corresponds.   
   
At the critical point the model is described by a minimal (non-diagonal) CFT   
with central charge $c=4/5$. It may be useful to briefly describe this model.   
It is indeed the simplest example of a non-diagonal minimal model of the   
Virasoro series~\cite{bpz}. However it can be also realized as the simplest   
{\sl diagonal} minimal model of the so called $W_3$ algebra~\cite{fz87}.   
  
Its operator content is composed by six primary fields, which however, due to the   
non-diagonal nature of the model, lead to a larger number of operators when the   
analytic and antianalytic sector are combined. Using the standard   
CFT notation $(r,s;r',s')$ which corresponds to the conformal dimensions    
$(h_{r,s},h_{r',s'})$ for the field, with $h_{r,s}$ given by 
\eq   
h_{r,s}=\frac{(6r-5s)^2-1}{120}   
\label{confweight}   
\en   
the most relevant operator in the energy sector is   
$\epsilon=(2,1;2,1)$ with scaling dimension $X_\epsilon=2h_{2,1}=4/5$,  
while the most relevant operators in the magnetic sector are the doublet $\sigma$ and $\bar\sigma$,  
of type $(3,3;3,3)$, with scaling dimension $X_\sigma=X_{\bar\sigma}=2h_{3,3}=2/15$.    
       
The nice feature of the $W_3$ description is that the operator content is   
composed by four fields only (besides the identity) and all the    
higher fields    
appear as secondary fields in the $W_3$ conformal families.

\section{Large distance expansion}   
\label{sect3}   
In this section we shall obtain the large distance behaviour of the two- and three-point functions of the model.   
As a preliminary step in this direction we shall first    
obtain an explicit expression for the two-particle form factors.   
\subsection{Form factors of order/disorder operators in three-state Potts model} \label{sect3.1}   
   
The continuum limit description of the model in which we are interested    
is given by the    
thermal perturbation of the above discussed CFT, i.e.   
\bea   
S=S_{\textrm{\tiny CFT}}+\tau \int \textrm{d}^2 x \; \epsilon(x)   
\label{Stau} 
\eea   
and belongs to the class of integrable QFTs \cite{Taniguchi}.    
   
In the computation of form factors it is fundamental to establish which is the   
nature of the basis of asymptotic states of the theory, i.e.\ which are the particle   
excitations which come into play. This is the main difference between the high-temperature and  
low-temperature phase of the model.    
   
At $\tau < 0$ there are three degenerate vacua and the excitations are the kinks $K_{j,j \pm 1}$, with $j=1,2,3$,  
which interpolate between the ground state $j$ and the ground state $j+1(\textrm{mod} 3)$. Form factors in this  
regime can be found in \cite{Delfino:1997ag}.   
   
At $\tau >0$ the ground state is unique, the ${\mathbb S}_3\simeq {\mathbb Z}_2 \times {\mathbb Z}_3$ symmetry is  
unbroken and its simplest realization is in terms of a doublet of particles $A$ and $\bar A$ transforming as   
\bea   
\Omega A = \omega A, \ \ \ \   \Omega \bar A = \omega^{-1} \bar A; \ \ \ \ \ \ \ \  \mathcal{C} A = \bar A   
\label{sym}   
\eea     
where $\Omega$ is the ${\mathbb Z}_3$ generator, $\mathcal{C}$ the charge conjugation operator and  
$\omega= e^{2 \pi i /3}$. The ${\mathbb Z}_3$ symmetry allows the existence of the fusion   
process    
\bea   
AA \to \bar A.   
\label{proc}   
\eea   
The two particle $S$-matrix for this integrable model turns out to be \cite{Zamolodchikov:zf,Koberle:sg}   
\bea   
| A(\theta_1)A(\theta_2) \rangle & = & u(\theta) | A(\theta_2)A(\theta_1) \rangle   
\nonumber \\   
| A(\theta_1) \bar A(\theta_2) \rangle & = & t(\theta) | A(\theta_2) \bar A(\theta_1) \rangle 
\eea   
where\footnote{The rapidity variables $\theta_i$ parameterize energy and momentum of the particles of mass $m$  
as $(p^0_i,p^1_i)=(m\cosh\theta_i,m\sinh\theta_i)$.} $\theta=\theta_1-\theta_2$ and    
\bea   
u(\theta) = t(i \pi - \theta) = \frac{\sinh\frac{1}{2}(\theta + \frac{2 \pi i}{3})}{\sinh\frac{1}{2}(\theta - \frac{2 \pi i}{3})}.   
\eea   
The pole present in the amplitude $u(\theta)$ located at $\theta= \frac{2 \pi i}{3}$ corresponds to the bound state (\ref{proc}).    
     
For the reasons discussed above, in the following we shall concentrate on the $\tau>0$ case.    
The symmetry of the model implies the presence of a doublet of spin operators $\sigma$ and $\bar \sigma$ which    
form a two-dimensional representation of ${\mathbb S}_3$ and transform as $A$ and $\bar A$ in (\ref{sym}).  
Duality also requires the existence of a doublet of disorder operators $\mu$, $\bar \mu$.    
   
The $n$-particle form factors of a given operator $\Phi$ are defined as   
\bea   
F_{a_1, \dots, a_n}^\Phi(\theta_1, \dots, \theta_n)=\langle 0 | \Phi(0)| A_{a_1} (\theta_1) \dots A_{a_n}  
(\theta_n) \rangle   
\eea   
with the indices $a_i$ referring to particle species.  
They satisfy a set of axioms which can be used to compute them explicitly   
\cite{KW,smibook, yuzam}.    
   
In the following we shall be  interested in the two-particle form factors for the $\sigma$ ($\bar \sigma$) and    
$\mu$ ($\bar \mu$) operators \cite{KS,Delfino:1997ag,BFK}. The spin operator satisfy (we recall that $\sigma$ and  
$\mu$ have non-zero form factors upon charged and neutral asymptotic states, respectively)   
\bea   
F_{A  A}^\sigma (\theta) = u(\theta) F_{A  A}^\sigma (\theta+2 \pi i )    
\eea   
with the residue condition due to the bound state pole of (\ref{proc})   
\bea   
-i \; \textrm{Res}_{\theta=\frac{2 \pi i}{3}} F_{A  A}^\sigma (\theta) = \Gamma^{\bar A}_{AA} F_{\bar A}^\sigma   
\eea   
where $F_{\bar A}^\sigma$ is the one-particle form factor and the three-particle coupling    
constant $\Gamma^{\bar A}_{AA}$ is given by   
\bea   
i \; \textrm{Res}_{\theta=\frac{2 \pi i}{3}} u (\theta) = (\Gamma^{\bar A}_{AA})^2 =\sqrt{3}.   
\eea   
Since the disorder operator $\mu$ and $\bar \mu$ are non-local with respect to the spin operators $\sigma$    
and $\bar \sigma$ which create the particles $A$ and $\bar A$, a phase factor $e^{\pm 2 \pi i/3}$ enters in    
the form factor equations   
\bea   
F_{A  \bar A}^\mu (\theta)= t(\theta)\; e^{- 2 \pi i/3} \; F_{A \bar A}^\mu (\theta+2 \pi i) \nonumber \\   
F_{\bar A A}^\mu (\theta)= t(\theta)\; e^{ 2 \pi i/3} \; F_{\bar A A}^\mu (\theta+2 \pi i).    
\eea   
Defining the vacuum expectation value of the disorder operator $\mu$ as $\langle \mu \rangle$, the residue conditions due to the kinematic pole follow   
\bea   
-i \; \textrm{Res}_{\theta=i \pi} F_{A  \bar A}^\mu (\theta) & = &\left(1-  e^{ 2 \pi i/3} \right) \langle \mu \rangle  \nonumber \\   
-i \; \textrm{Res}_{\theta=i \pi} F_{\bar A A}^\mu (\theta) & = &\left(1-  e^{ -2 \pi i/3} \right) \langle \mu \rangle.   
\eea   
In order to work out the solutions, we introduce integral representations for both $u(\theta)$ and $t(\theta)$. Let us define   
\bea   
f_\alpha(\theta) & = & - \frac{\sinh\frac{1}{2}(\theta + i \pi \alpha)}{\sinh\frac{1}{2}(\theta - i \pi \alpha)} = \nonumber \\   
& = &  \exp \left[ 2 \int_0^\infty \frac{\textrm{d}x}{x} \frac{\sinh(1-\alpha)x}{\sinh x} \sinh \frac{\theta x}{i \pi} \right]\,\,.   
\eea   
Then, we have immediately   
\bea   
u(\theta)= - f_{2/3}(\theta), \ \ \ \ \ \   t(\theta)= f_{1/3}(\theta).   
\eea   
Since the solution to the functional equation $F_\alpha (\theta) = f_\alpha (\theta)F_\alpha (\theta+2 \pi i)$ admits the integral  
representation   
\bea   
&& F_\alpha (\theta) = \exp \left[ 2 \int_0^\infty \frac{\textrm{d}x}{x} \frac{\sinh(1-\alpha)x}{\sinh^2 x} \sin^2 \frac{(i \pi -\theta) x}{2 \pi} \right], \\   
&& F_\alpha (\theta) \sim \exp \frac{1-\alpha}{2} |\theta|\,, \ \ \ \ \ |\theta| \to \infty   
\eea   
one finds the desired expressions for the 2-particle form factors   
\bea   
F_{A  A}^\sigma (\theta) & = & \Gamma^{\bar A}_{AA} \; F_{\bar A}^\sigma \;  \frac{i \sinh \theta/2}{\sinh\frac{1}{2}(\theta + \frac{2 \pi i}{3})\; \sinh\frac{1}{2}(\theta - \frac{2 \pi i}{3})} \frac{F_{2/3}(\theta)}{2 F_{2/3}(2 \pi i /3)} \nonumber \\   
F_{A  \bar A}^\mu (\theta) & = & - \langle \mu \rangle \; \frac{\sqrt{3}}{2} \frac{e^{-\theta/6}}{\cosh \theta/2} \; F_{1/3}(\theta) \nonumber \\   
F_{\bar A A}^\mu (\theta) & = & - \langle \mu \rangle \; \frac{\sqrt{3}}{2} \frac{e^{\theta/6}}{\cosh \theta/2} \; F_{1/3}(\theta).   
\eea   
The symmetry of the model requires that the remaining form factors for the conjugated operators are given by    
\bea   
F_{\bar A  \bar A}^{\bar \sigma} (\theta) = F_{A  A}^\sigma (\theta); \ \ \ \ \ \ F_{\bar A  \bar A}^\sigma (\theta) = F_{A  A}^{\bar \sigma} (\theta)=0   
\eea   
\bea   
F_{A  \bar A}^{\bar \mu} (\theta) = F_{\bar A  A}^\mu (\theta); \ \ \ \ \ \ F_{\bar A   A}^{\bar \mu} (\theta) = F_{A  \bar A}^{ \mu} (\theta).   
\label{ff} 
\eea   
The relative normalizations between order and disorder form factors can be obtained by means of the cluster condition \cite{cds,Delfino:1997ag}   
\bea   
\lim_{\theta \to \infty}  |  F_{\bar A A}^\mu (\theta)  | = \frac{F_{\bar A}^\sigma F_{A}^{\bar \sigma}}{\langle \mu \rangle }   
\label{cluster} 
\eea   
(symmetry requires $F_{\bar A}^\sigma =F_{A}^{\bar \sigma}$).   
 
The knowledge of the first few form factors gives access to approximate expressions for correlation functions. Their explicit expressions for the  operators $\sigma$, $\bar \sigma$, $\mu$ and $\bar \mu$ are listed below   
\bea   
\langle \sigma (x) \bar \sigma(0) \rangle &  =  & \frac{F_{\bar A}^\sigma F_{A}^{\bar \sigma}}{\pi} K_0(m |x|) \; + \; \frac{1}{\pi^2} \int_0^\infty \textrm{d} \theta \; F_{A  A}^\sigma (2 \theta) F_{A  A}^\sigma (-2 \theta) \; K_0(2 m |x| \cosh \theta) + \dots \label{corrs}    
\\   
\langle \mu (x) \bar \mu(0) \rangle &  =  &  \langle \mu \rangle^2  \; +  \; \frac{1}{\pi^2} \int_0^\infty \textrm{d} \theta \; \left[ F_{A  \bar A}^{\mu} (2 \theta)  F_{  \bar A A}^{\bar \mu} (-2 \theta) +  F_{\bar A  A}^{\mu} (2 \theta)  F_{A \bar A}^{\bar \mu} (-2 \theta) \right]\; K_0(2 m |x| \cosh \theta) + \dots   
\nonumber \\   
\langle \mu (x)  \mu (0) \rangle &  =  & \langle \mu \rangle^2  \; +  \; \frac{1}{\pi^2} \int_0^\infty \textrm{d} \theta \; \left[ F_{A  \bar A}^{\mu} (2 \theta)  F_{  \bar A A}^{\mu} (-2 \theta) +  F_{\bar A  A}^{\mu} (2 \theta)  F_{A \bar A}^{\mu} (-2 \theta) \right]\; K_0(2 m |x| \cosh \theta) + \dots \nonumber   
\eea    
where ellipsis stand for terms of order $e^{-3 m |x|}$ for large $|x|$. We remember that symmetry requires   
\bea   
& & \langle \sigma (x) \bar \sigma(0) \rangle = \langle \bar \sigma (x)  \sigma(0) \rangle, \nonumber \\   
& & \langle \mu (x) \bar \mu(0) \rangle = \langle \bar \mu (x)  \mu(0) \rangle , \nonumber \\   
& & \langle \mu (x) \mu(0) \rangle = \langle \bar \mu (x)  \bar \mu(0) \rangle . \nonumber    
\eea

\subsection{Three-point correlation function}   
\label{sect3.2}   
As first noticed in \cite{Chim:tf}, the pole due to the process (\ref{proc}) has a peculiar effect on the spectral expansion of the three-point correlation function $G^{(3)}(x_1,x_2,x_3)=\langle \sigma(x_1) \sigma(x_2) \sigma(x_3) \rangle$. We recall that (in the notations of high-temperature phase) it takes the following form   
\bea   
\label{trefu}   
G^{(3)}(x_1,x_2,x_3) \ = \ \frac{(F_{\bar A}^\sigma)^3 \; \Gamma^{\bar A}_{AA}}{\pi} \ K_0(m r_Y)+O(e^{-m \rho})   
\eea   
where it is assumed that all the angles of the triangle $(x_1,x_2,x_3)$ are less than $2 \pi /3$ and   
$r_Y$ denotes the minimal total length of lines connecting the 3 spins and is given by   
\bea   
r_Y \ = \ r_1 + r_2 + r_3 = \sqrt{\frac{1}{2} R_{12}^2+\frac{1}{2} R_{23}^2+\frac{1}{2} R_{31}^2+  
2 \sqrt{3} S_\bigtriangleup}\,\,. 
\label{giusta}   
\eea   
Here $S_\bigtriangleup$ is the area of the triangle $(x_1,x_2,x_3)$ and the meaning of $r_i$ and $R_{jk}$ is    
clear from figure \ref{f1}.    
The point at which the three segments of lengths $r_i$ join is usually called ``Steiner point''.   
If one of the angles becomes larger than $2 \pi /3$, then the large distance asymptotic is dominated by $O (e^{-m \rho})$ where   
\bea   
\rho = \min(R_{12}+R_{23},R_{23}+R_{31},R_{12}+R_{31}).   
\label{mini}   
\eea   
The aim of this section is to give a proof of the previous result together with an explicit expression of the subleading term $O (e^{-m \rho})$.    
   
The spectral expansion of the correlation function $G^{(3)}$ can be obtained inserting the resolution of the identity between the operators which appear in the   
correlator. More explicitly, we can write the general expansion as   
\bea   
G^{(3)}(x_1,x_2,x_3) \ = \ \sum_{k,l}     
\langle 0 | \sigma (x_1) | k \rangle \langle k | \sigma (x_2) | l \rangle \langle l | \sigma (x_3) | 0 \rangle = \sum_{k,l}  G^{(3)}_{[k,l]}(x_1,x_2,x_3)   
\eea   
and hence we will use the notation $[k,l]$ to identify a given contribution 
($k$ and $l$ denote the number of particles present in the states $|k\rangle$ and $|l\rangle$, respectively).   
   
It is understood that all the following formul\ae~ are valid in the limit of \emph{large} triangles, namely when   
the condition $m R_{12}$, $m R_{23}$, $m R_{13} \gg 1$ is fulfilled.

\subsection{$[1,1]$ contribution}   
\begin{figure}   
\begin{center}   
\includegraphics[width=9cm]{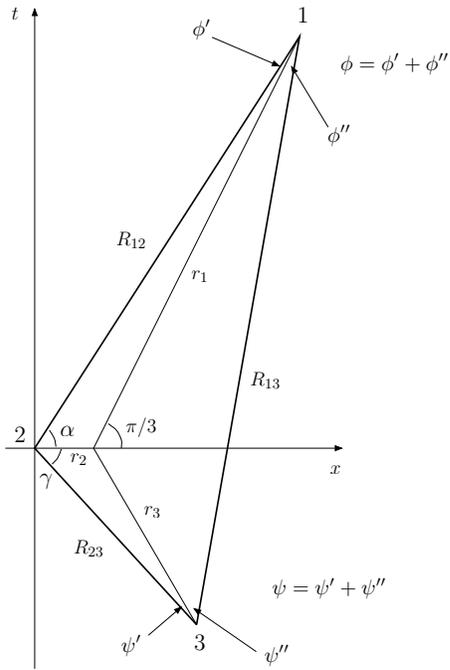}   
\caption{Triangle associated to the three-point correlation function. The orientation of the triangle has been   
chosen so that one of the three segments joining the vertices with the Steiner point lies on the real   
axis.}   
\label{f1}   
\end{center}   
\end{figure}

The leading term of the spectral series is 
\bea   
G^{(3)}_{[1,1]}(x_1,x_2,x_3) \ = \  \int_{-\infty}^{\infty} \frac{\textrm{d} \beta}{2 \pi}  \int_{-\infty}^{\infty} \frac{\textrm{d} \theta}{2 \pi}    
\langle 0 | \sigma (x_1) | \bar A(\beta) \rangle \langle \bar  A(\beta) | \sigma (x_2) | A (\theta) \rangle \langle A (\theta) | \sigma (x_3) | 0 \rangle\,.   
\nonumber   
\eea   
Extracting the space-time dependence of the matrix elements the previous expression becomes  
(see figure \ref{f1} for conventions)   
\bea   
(F_{\bar A}^\sigma)^2 \int_{-\infty}^{\infty} \frac{\textrm{d} \beta}{2 \pi}  \int_{-\infty}^{\infty} \frac{\textrm{d} \theta}{2 \pi}    
F_{A  A}^{\sigma} (\beta - \theta + i \pi) \exp \left[ - i m \left( R_{12} \sinh (\beta -i \alpha ) -    
R_{23} \sinh (\theta +i \gamma ) \right) \right] .      
\label{init}   
\eea   
Then, taking into account that the residue on the bound state gives the condition   
\bea   
-i \; \textrm{Res}_{\theta_-=\frac{2 \pi i}{3}-i\alpha_+} F_{A A}^{\sigma} (\theta_- + i \alpha_+) = \Gamma^{\bar A}_{AA} F_{\bar A}^\sigma 
\nonumber   
\eea   
($\alpha_+=\alpha+\gamma$) and using the Cauchy's theorem (see Appendix A for details), we can write $G^{(3)}_{[1,1]}$ as follows   
\bea   
&& G^{(3)}_{[1,1]}(x_1,x_2,x_3) = H \left( \frac{2 \pi}{3}-\alpha_+ \right) \; \frac{(F_{\bar A}^\sigma)^3 \; \Gamma^{\bar A}_{AA} }{ \pi} K_0(m r_Y)+  \label{G11}   \\   
&& \hspace{-0.5cm} +  (F_{\bar A}^\sigma)^2\int_{-\infty}^{\infty} \frac{\textrm{d} \beta}{2 \pi}  \int_{-\infty}^{\infty} \frac{\textrm{d} \theta}{2 \pi}    
F_{ A  A}^{ \sigma} (\beta - \theta + i (\alpha + \gamma)) \exp \left[ - m \left( R_{12} \cosh \beta +    
R_{23} \cosh \theta  \right) \right]. \nonumber   
\eea    
where $H(z)$ is the usual step-function and $r_Y=\sum_{i=1}^3 r_i$ (see fig. \ref{f1})

\subsection{$[1,2]$ \& $[2,1]$ contributions}    
According to the general formula, the next-to-leading term of the spectral series is given by    
\bea   
G^{(3)}_{2p} (x_1,x_2,x_3) \ = \   && \hspace{-0.5cm}    
G^{(3)}_{[1,2]} (x_1,x_2,x_3)+G^{(3)}_{[2,1]} (x_1,x_2,x_3) = \nonumber\\   
&& \hspace{-2.5cm} = \frac{1}{2!} \int_{-\infty}^{\infty} \frac{\textrm{d} \beta}{2 \pi}  \int_{-\infty}^{\infty} \frac{\textrm{d} \theta_1 \textrm{d} \theta_2}{(2 \pi)^2}    
\langle 0 | \sigma (x_1) | \bar A(\beta) \rangle \langle {\bar A}(\beta) | \sigma (x_2) | {\bar A} (\theta_1) {\bar A} (\theta_2)\rangle \langle {\bar A} (\theta_2) {\bar A} (\theta_1) | \sigma (x_3) | 0 \rangle+   
\nonumber \\   
&& \hspace{-2.5cm}+\frac{1}{2!} \int_{-\infty}^{\infty} \frac{\textrm{d} \beta}{2 \pi}  \int_{-\infty}^{\infty} \frac{\textrm{d} \theta_1 \textrm{d} \theta_2}{(2 \pi)^2}     
\langle 0 | \sigma (x_1) |A (\theta_1)  A (\theta_2) \rangle \langle A (\theta_2)  A (\theta_1) | \sigma (x_2) |A(\beta)\rangle \langle A(\beta) | \sigma (x_3) | 0 \rangle \nonumber   
\eea   
and hence, extracting the space-time dependence of the matrix elements, we obtain    
\bea   
G^{(3)}_{2p} (x_1,x_2,x_3) \ = \   && \hspace{-0.5cm}    
G^{(3)}_{[1,2]} (x_1,x_2,x_3)+G^{(3)}_{[2,1]} (x_1,x_2,x_3) = \nonumber \\   
&& \hspace{-2.5cm}= \frac{1}{2!} \int_{-\infty}^{\infty} \frac{\textrm{d} \beta}{2 \pi}  \int_{-\infty}^{\infty} \frac{\textrm{d} \theta_1 \textrm{d} \theta_2}{(2 \pi)^2}    
\langle 0 | \sigma (0) | \bar A(\beta) \rangle \langle {\bar A}(\beta) | \sigma (0) | {\bar A} (\theta_1) {\bar A} (\theta_2)\rangle \langle {\bar A} (\theta_2) {\bar A} (\theta_1) | \sigma (0) | 0 \rangle \times  
 \nonumber \\   
&& \hspace{0.3cm}\exp \left[ - i m \left( R_{12} \sinh (\beta -i \alpha ) -    
R_{23} (\sinh (\theta_1 +i \gamma )+\sinh (\theta_2 +i \gamma )) \right) \right] +   
\nonumber \\   
&& \hspace{-2.5cm}+\frac{1}{2!} \int_{-\infty}^{\infty} \frac{\textrm{d} \beta}{2 \pi}  \int_{-\infty}^{\infty} \frac{\textrm{d} \theta_1 \textrm{d} \theta_2}{(2 \pi)^2}     
\langle 0 | \sigma (0) |A (\theta_1)  A (\theta_2) \rangle \langle A (\theta_2)  A (\theta_1) | \sigma (0) |A(\beta)\rangle \langle A(\beta) | \sigma (0) | 0 \rangle \times \nonumber \\   
&& \hspace{-0.5cm}\exp \left[ - i m \left( R_{12}( \sinh ( \theta_1-i \alpha )+\sinh ( \theta_2-i \alpha ) )-    
R_{23} \sinh (\beta +i \gamma ) \right) \right].     
\label{g32p}   
\eea   
It turns out that the above term  gives a contribution to the correlator of the same order of magnitude of    
$G^{(3)}_{[1,1]}$. In order to understand this let us  first    
notice that the matrix elements have to be rewritten in terms of   
form factors (which are the quantities that one actually computes in integrable field theories).    
Such a reduction to form factors is achieved by iterating the crossing relation \cite{KW,smibook,DMS}, and in the    
case of $G^{(3)}_{[1,2]}$ (the treatment of $G^{(3)}_{[2,1]}$ will follow the same guidelines) we have   
\bea   
\langle {\bar A}(\beta) | \sigma (0) | {\bar A} (\theta_1) {\bar A} (\theta_2)\rangle =  \   && \hspace{-0.5cm} \langle 0 | \sigma (0) | A (\beta + i \pi) {\bar A} (\theta_1) {\bar A} (\theta_2)\rangle + \\   
&& \hspace{-1.5cm} + 2 \pi \delta (\beta -\theta_1) \langle 0 | \sigma (0) | \bar A(\beta)   
\rangle + 2 \pi \delta (\beta -\theta_2) u(\theta_1-\beta) \langle 0 | \sigma (0) | \bar A(\beta) \rangle =   
\nonumber \\   
&& \hspace{-1.5cm} = F^{\sigma}_{A {\bar A} {\bar A} }(\beta + i \pi,\theta_1,\theta_2) +    
2 \pi ( \delta (\beta -\theta_1) + \delta (\beta -\theta_2) u(\theta_1-\beta) ) F^{\sigma}_{ {\bar A} }\,.  
\nonumber    
\eea   
Plugging the previous expression in $G^{(3)}_{[1,2]}$ and performing the integration over the   
delta functions, we obtain   
\bea   
G^{(3)}_{[1,2]} (x_1,x_2,x_3) =   && \hspace{-0.5cm} \frac{F^{\sigma}_{\bar A}}{2!} \int_{-\infty}^{\infty} \frac{\textrm{d} \beta}{2 \pi}  \int_{-\infty}^{\infty} \frac{\textrm{d} \theta_1 \textrm{d} \theta_2}{(2 \pi)^2}   
F^{\sigma}_{A {\bar A} {\bar A} }(\beta + i \pi,\theta_1,\theta_2)   
F^{\sigma}_{A  A} (\theta_2-\theta_1) \times \\   
&& \hspace{0.3cm}\exp \left[ - i m \left( R_{12} \sinh (\beta -i \alpha ) -    
R_{23} (\sinh (\theta_1 +i \gamma )+\sinh (\theta_2 +i \gamma )) \right) \right] +   
\nonumber \\   
&& \hspace{-3.5cm}+\frac{(F^{\sigma}_{\bar A})^2}{2!}   \int_{-\infty}^{\infty} \frac{\textrm{d} \theta_1 \textrm{d} \theta_2}{(2 \pi)^2}     
F^{\sigma}_{A  A} (\theta_2-\theta_1)   
\exp \left[ - i m \left( R_{12} \sinh (\theta_1 -i \alpha ) -    
R_{23} (\sinh (\theta_1 +i \gamma )+\sinh (\theta_2 +i \gamma ))  \right) \right] +   
\nonumber \\   
&& \hspace{-3.5cm}+\frac{(F^{\sigma}_{\bar A})^2}{2!}   \int_{-\infty}^{\infty} \frac{\textrm{d} \theta_1 \textrm{d} \theta_2}{(2 \pi)^2}     
F^{\sigma}_{A  A} (\theta_1-\theta_2)   
\exp \left[ - i m \left( R_{12} \sinh (\theta_2 -i \alpha ) -    
R_{23} (\sinh (\theta_1 +i \gamma )+\sinh (\theta_2 +i \gamma ))  \right) \right] \nonumber    
\eea   
where in the last line we have used the fact that $F^{\sigma}_{A  A} (\theta)= u(\theta)F^{\sigma}_{A  A} (-\theta) $.    
A brief inspection of such an expression shows that the last two terms coincide (to see this one can make the exchange $\theta_1 \leftrightarrow \theta_2$ in one of them), and hence we can write   
\bea   
G^{(3)}_{[1,2]} (x_1,x_2,x_3) =   && \hspace{-0.5cm} \frac{F^{\sigma}_{\bar A}}{2!} \int_{-\infty}^{\infty} \frac{\textrm{d} \beta}{2 \pi}  \int_{-\infty}^{\infty} \frac{\textrm{d} \theta_1 \textrm{d} \theta_2}{(2 \pi)^2}   
F^{\sigma}_{A {\bar A} {\bar A} }(\beta + i \pi,\theta_1,\theta_2)   
F^{\sigma}_{A  A} (\theta_2-\theta_1) \times \\   
&& \hspace{0.3cm}\exp \left[ - i m \left( R_{12} \sinh (\beta -i \alpha ) -    
R_{23} (\sinh (\theta_1 +i \gamma )+\sinh (\theta_2 +i \gamma )) \right) \right] +   
\nonumber \\   
&& \hspace{-3.5cm}+\ (F^{\sigma}_{\bar A})^2\int_{-\infty}^{\infty} \frac{\textrm{d} \theta_1 \textrm{d} \theta_2}{(2 \pi)^2}     
F^{\sigma}_{A  A} (\theta_2-\theta_1)\exp \left[ - i m \left( R_{12} \sinh (\theta_1 -i \alpha ) -    
R_{23} (\sinh (\theta_1 +i \gamma )+\sinh (\theta_2 +i \gamma ))  \right) \right].    
\nonumber     
\eea    
Such an expression for $G^{(3)}_{[1,2]}$ shows that the disconnected parts sum up giving a contribution which is quite similar to $G^{(3)}_{[1,1]}$. Hence, if we pose   
\bea   
G^{(3),disc}_{[1,2]} (x_1,x_2,x_3) =   && \hspace{-0.5cm}\ (F^{\sigma}_{\bar A})^2 \,   \int_{-\infty}^{\infty} \frac{\textrm{d} \theta_1 \textrm{d} \theta_2}{(2 \pi)^2}     
F^{\sigma}_{A  A} (\theta_2-\theta_1)  \times \\   
&& \hspace{0.3cm}\exp \left[ - i m \left( R_{12} \sinh (\theta_1 -i \alpha ) -    
R_{23} (\sinh (\theta_1 +i \gamma )+\sinh (\theta_2 +i \gamma ))  \right) \right]   
\nonumber     
\eea    
and we use the Cauchy's theorem taking into account the residue condition on the bound state, we obtain  
(see Appendix A for details)   
\bea   
G^{(3),disc}_{[1,2]} (x_1,x_2,x_3) \ = \ && \hspace{-0.5cm} - H \left(\psi - \frac{2 \pi}{3} \right) \;  
\frac{(F_{\bar A}^\sigma)^3 \; \Gamma^{\bar A}_{AA} }{ \pi} K_0(m r_Y)+   \label{G12}     \\   
&& \hspace{-0.5cm} +  (F_{\bar A}^\sigma)^2\int_{-\infty}^{\infty} \frac{\textrm{d} \theta_1 \textrm{d} \theta_2}{(2 \pi)^2}  
F_{ A  A}^{ \sigma} (\theta_2 - \theta_1 + i \psi) \exp \left[ - m \left( R_{23} \cosh \theta_2 +    
R_{13} \cosh \theta_1  \right) \right]. \nonumber  
\eea    
One can notice the change of sign in front of the function $H(z)$ and the difference in its argument with   
respect to $G^{(3)}_{[1,1]}$.   
   
As anticipated, we can treat $G^{(3)}_{[2,1]}$ in exactly the same way as $G^{(3)}_{[1,2]}$. At the end of the computation  
we are left with the following expression for $G^{(3),disc}_{[2,1]}$   
\bea   
G^{(3),disc}_{[2,1]} (x_1,x_2,x_3)  \ = \   && \hspace{-0.5cm} - H \left(\phi - \frac{2 \pi}{3} \right) \;  
\frac{(F_{\bar A}^\sigma)^3 \; \Gamma^{\bar A}_{AA} }{ \pi} K_0(m r_Y)+  \label{G21}     \\   
&& \hspace{-0.5cm} +  (F_{\bar A}^\sigma)^2\int_{-\infty}^{\infty} \frac{\textrm{d} \theta_1 \textrm{d} \theta_2}{(2 \pi)^2}  
F_{ A  A}^{ \sigma} (\theta_1 - \theta_2 + i \phi) \exp \left[ - m \left( R_{13} \cosh \theta_1 +    
R_{12} \cosh \theta_2  \right) \right].   \nonumber 
\eea    
   
Collecting all the previous contributions we can write down the explicit expression of the spectral series for the three-point function up to one-particle contributions   
\bea   
G^{(3)} (x_1,x_2,x_3)  \ = \   && \hspace{-0.7cm} G^{(3)}_{[1,1]}+G^{(3),disc}_{[1,2]}+   
G^{(3),disc}_{[2,1]} + \dots =    
\nonumber \\   
&& \hspace{-3.5cm} =   
\left\{ H \left(\frac{2 \pi}{3} -  \alpha_+ \right) - H \left(\psi - \frac{2 \pi}{3} \right)   
 - H \left(\phi - \frac{2 \pi}{3} \right) \right\} \; \frac{(F_{\bar A}^\sigma)^3 \; \Gamma^{\bar A}_{AA} }{ \pi} K_0(m r_Y)+      
\nonumber  \\   
&& \hspace{-3.5cm} +  (F_{\bar A}^\sigma)^2\int_{-\infty}^{\infty} \frac{\textrm{d} \beta \textrm{d} \theta}{(2 \pi)^2}    
F_{ A  A}^{ \sigma} (\beta - \theta + i \alpha_+) \exp \left[ - m \left( R_{12} \cosh \beta +    
R_{23} \cosh \theta  \right) \right] +   
\nonumber \\   
&& \hspace{-0.7cm}+ F_{ A  A}^{ \sigma} (\beta - \theta + i \psi) \exp \left[ - m \left( R_{23} \cosh \beta +    
R_{13} \cosh \theta  \right) \right] +   
\nonumber \\   
&& \hspace{-0.7cm} +F_{ A  A}^{ \sigma} (\beta - \theta + i \phi) \exp \left[ - m \left( R_{13} \cosh \beta +    
R_{12} \cosh \theta  \right) \right]+ \dots.      
\label{g3symm}   
\eea    
Some comments about the previous expression are in order.    
\begin{itemize}   
\item{The combination of $H(z)$ functions which appears in the first term   
nicely realizes the condition stated at the beginning of the section. It is not difficult to show that    
only two possibilities are allowed: either all the angles are less than $2 \pi/3$ and    
such a prefactor is 1, or one of them is larger than $2 \pi /3$ and then the prefactor is zero.   
}   
\item{The second term of (\ref{g3symm}) shows explicitly that its asymptotic behaviour for large    
distances is given by    
$O (e^{-m \rho})$ where   
\bea   
\rho = \min(R_{12}+R_{23},R_{23}+R_{31},R_{12}+R_{31}).   
\nonumber   
\eea   
}   
\end{itemize}   
\subsection{Isosceles \& Equilateral triangles}    
Formula (\ref{g3symm}) can be simplified by means of a wise choice of the geometry of the   
triangle.    
   
Let us begin with the case of isosceles triangles. We decide to make the choice: $R=R_{12}=R_{23}$,   
$S=R_{13}=2R \sin \alpha$, and $\alpha=\gamma$ ($\alpha_+=2 \alpha$), $\psi = \phi = \pi/2 - \alpha$ (it is easy to   
show that any other choice of the isosceles triangle will be equivalent to the previous one). Hence   
(\ref{g3symm}) will take the following form   
\bea   
G^{(3)} (x_1,x_2,x_3)  \ \simeq \   && \hspace{-0.7cm}    
H \left(\frac{ \pi}{3} - \alpha \right) \; \frac{(F_{\bar A}^\sigma)^3 \; \Gamma^{\bar A}_{AA} }{ \pi} K_0   
 \left( 2 m R    
\cos \left( \frac{ \pi}{3} - \alpha \right) \right)+      
\nonumber  \\   
&& \hspace{-3.5cm} + 2 (F_{\bar A}^\sigma)^2 \int_{-\infty}^{\infty} \frac{\textrm{d} \theta}{(2 \pi)^2}    
F_{ A  A}^{ \sigma} ( \theta + 2 i \alpha) K_0 \left( 2 m R \cosh \frac{\theta}{2}   
\right) + \nonumber \\   
&& \hspace{-3.5cm} + 2 (F_{\bar A}^\sigma)^2\int_{-\infty}^{\infty} \frac{\textrm{d} \beta \textrm{d} \theta}{(2 \pi)^2}    
F_{ A  A}^{ \sigma} (\beta - \theta + i (\pi/2 - \alpha)) \exp \left[ - m R \left(  \cosh \beta +    
2 \sin \alpha \cosh \theta  \right) \right]\,.     
\nonumber   
\eea    
An even simpler form can be found when the case of equilateral triangles is considered. Such a choice corresponds to   
fixing $\alpha= \pi/6$ in the previous formula (obviously one can obtain the same result directly from    
formula (\ref{g3symm})) finding   
\bea   
G^{(3)} (x_1,x_2,x_3)  \ \simeq \   && \hspace{-0.7cm}    
\frac{(F_{\bar A}^\sigma)^3 \; \Gamma^{\bar A}_{AA} }{ \pi} K_0   
\left(\sqrt{3}\,  m R  \right)+      
\nonumber  \\   
&& \hspace{-0.9cm} + \    
6 (F_{\bar A}^\sigma)^2 \int_{-\infty}^{\infty} \frac{\textrm{d} \theta}{(2 \pi)^2}    
F_{ A  A}^{ \sigma} ( \theta +  i \pi/ 3) K_0 \left( 2 m R \cosh \frac{\theta}{2}   
\right).   
\label{g3equi}   
\eea    
    
\section{Monte-Carlo simulations}   
\label{sect4}   
  
\begin{figure}[tbp]  
\centerline{\includegraphics[height=30mm]{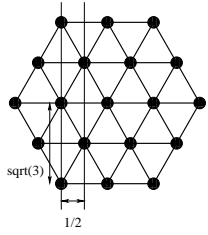}}   
\caption{Triangular lattice.} 
\label{fig:triang}    
\end{figure}

We performed Monte-Carlo simulations of the 3-state Potts model on a 
triangular lattice (where each site has six nearest neighbors, see 
fig.~\ref{fig:triang}) with honeycomb boundary conditions.  This 
lattice has a larger symmetry than a square lattice and allowed us to 
study 3-point functions with three exactly equidistant points.  A 
Swendsen-Wang cluster algorithm was used and 2- and 3-point functions 
were computed from improved estimators.  The latter proved essential 
for the 3-point function, which could not have been measured to as 
large distances with conventional estimators.  Simulations were 
performed at three different couplings, see table~\ref{tab:sim} below. 
In the table, a volume $60\times180$, for instance, refers to a 
hexagon with side $60$. 
    
\subsection{Determination of the mass and $F^{\sigma}_{\bar A}$}   
\label{sect4.1}   
The long-distance expansion for the 2-point function has two free   
parameters, the mass (inverse correlation length) and the   
1-particle form factor.  For each lattice, the mass was extracted from   
the effective mass, defined in terms of the wall-wall correlator   
$G^\ell_0(x)$ as\footnote{We denote lattice quantities which differ 
  in normalization from their field theory counterparts by an index 
  $\ell$.} 
\begin{equation}\label{eq:meff}   
  m_{\text{eff}}(x+\tfrac14) \equiv 2 \ln\frac {G^\ell_0(x)} {G^\ell_0(x+\tfrac12)}   
\end{equation}   
On a triangular lattice, 
\begin{equation}   
  G^\ell_0(x) = \sum_y \langle \sigma_\ell(x,y) \bar\sigma_\ell(0,0) \rangle   
        \simeq \frac{1}{\sqrt{3}} \int \dd y\,   
        \langle \sigma_\ell(x,y) \bar\sigma_\ell(0,0) \rangle   
\end{equation}   
where the factor $1/\sqrt{3}$ comes from the distance between points along   
the wall on a triangular lattice, see fig.~\ref{fig:triang}.   
Note also that a displacement of $\tfrac12$ instead of $1$ is possible   
on such a lattice.   
   
\begin{figure}[htbp]   
  \centering   
  \includegraphics[width=\epswid]{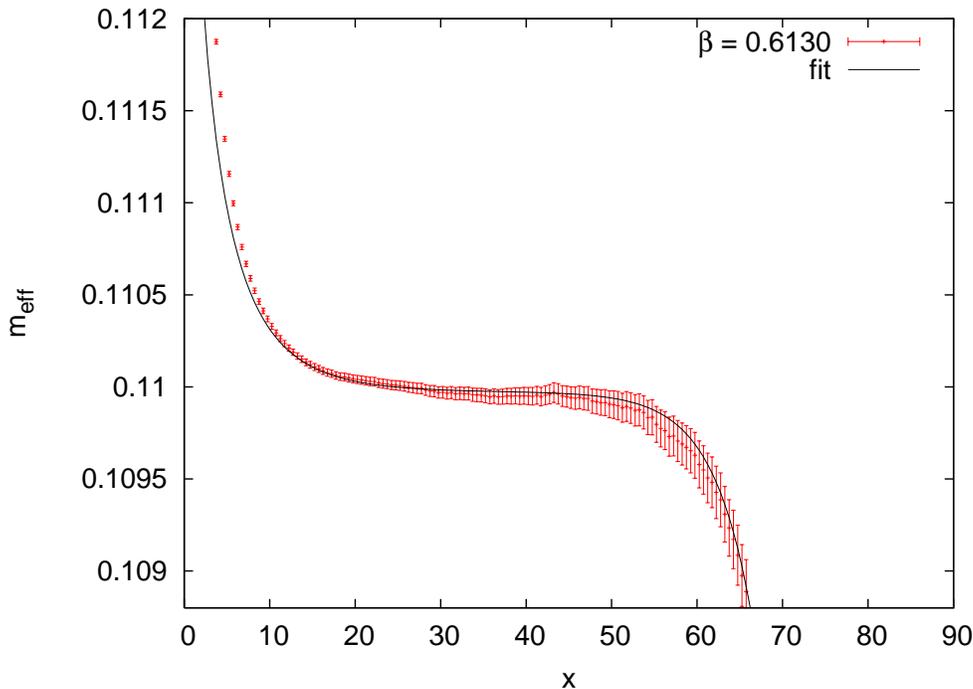}   
  \caption{Effective mass at $\beta=0.613$ and 
    1-parameter fit from eq.\ (\ref{eq:meff2}) including the 
    long-distance expansion up to 2 particles and finite-size 
    effects.  $x$ has period 180.} 
  \label{fig:effmass}   
\end{figure}

The prediction of 1- and 2-particle contributions for the   
propagator including finite-size effects for the 1-particle   
contribution (those from the 2-particle contribution are negligible) is   
\begin{equation}\label{eq:G0-2p}   
  G^\ell_0(x) = \frac {(F_{\ell})^2} {\sqrt{3} m} 
  \left( \ee^{-m x} + g_{2}(m x) + \ee^{-m(L-x)} \right)   
\end{equation}   
where 
\begin{equation}   
  g_{2}(m x)   
  = \frac{1}{2\pi} \int_0^\infty \dd\theta    
  \left| \frac{ F^\sigma_{A A}(2\theta) }{ F^\sigma_{\bar A} } \right|^2    
  \frac{\exp[-2m x \cosh\theta]}{\cosh\theta}   
\end{equation}   
and $F_\ell$ is the 1-particle form factor $F^\sigma_{\bar A}$ in lattice normalization. 
The corresponding effective mass is   
\begin{equation}\label{eq:meff2}   
  m_{\text{eff}}(x+\tfrac14) = m +    
  \delta m_{\text{eff}}(x+\tfrac14,m,L)   
\end{equation}   
with   
\begin{equation}\label{eq:meff-corr}   
  \delta m_{\text{eff}}(x+\tfrac14,m,L) = 2 \ln \frac   
  { 1 + \ee^{m x} g_2(m x) + \ee^{-m(L-2 x)} }   
  { 1 + \ee^{m(x+1/2)} g_2(m(x+\tfrac12)) + \ee^{-m(L-2 x-1)} }   
\end{equation}   
The fit with this function is performed by subtracting $\delta 
m_{\text{eff}}(x,m,L)$ with some trial $m$ from the effective mass 
data, fitting the result to a constant, and then iteratively improving 
$m$.  This procedure is very stable because the corrections are small 
in the range of the fit.  The fit at $\beta=0.613$ is 
shown in fig.~\ref{fig:effmass}.  It confirms the analytic prediction 
very well.  The resulting masses and correlation lengths, together 
with the fit ranges used, can be found in table~\ref{tab:sim}.  The 
error estimates were obtained with a jackknife analysis. 
  
\begin{table} 
  \centering  
  \begin{tabular}{cccccccc} 
    $\beta$     & volume                & $\xi$ & $m$   & $F_{\ell}$    & $x_{\text{min}}$      & $x_{\text{max}}$      & $\Gamma^{\bar A}_{A A}$ \\\hline 
    0.5900      & $60\times180$         & $4.5184(9)$   & $0.22132(5)$  & $0.92179(19)$ & $7.5$ & $25$  & $1.2609(16)$ \\ 
    0.6130      & $120\times360$        & $9.093(3)$    & $0.10997(3)$  & $0.8233(2)$   & $15$  & $50$  & $1.287(3)$ \\ 
    0.6231      & $240\times720$        & $18.232(7)$   & $0.054849(19)$        & $0.7420(3)$   & $25$  & $125$ & $1.299(3)$ 
  \end{tabular} 
  \caption{Simulation parameters.} 
  \label{tab:sim} 
\end{table} 
 
\begin{figure}[htbp]   
  \centering   
  \includegraphics[width=1\epswid]{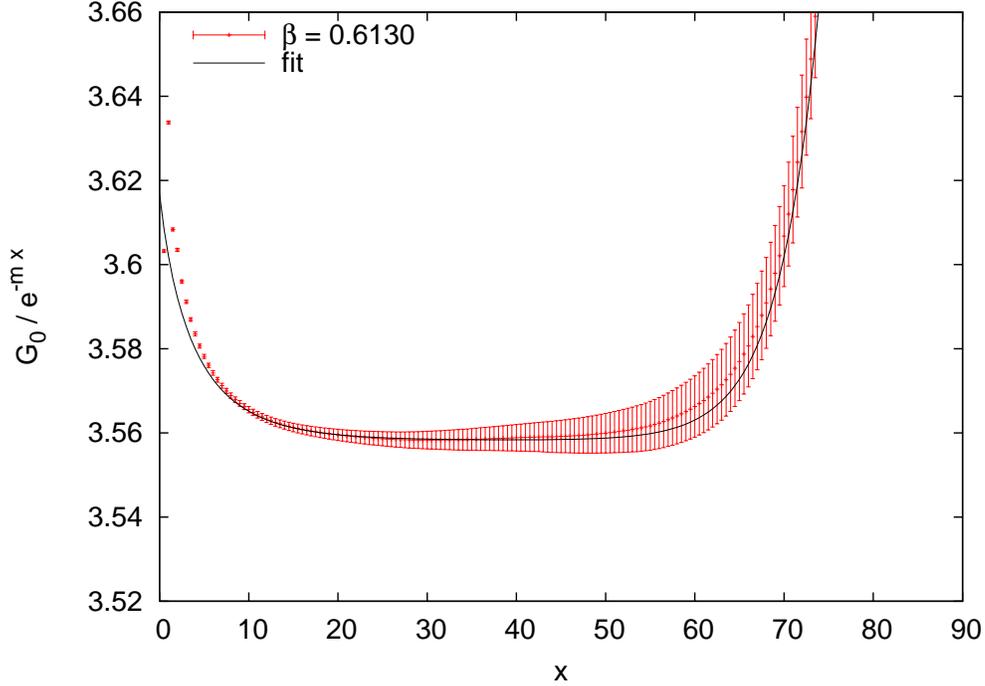}   
  \caption{Wall-wall correlator at $\beta=0.613$ and   
    2-particle prediction with finite-size effects.  Only the prefactor has 
    been fitted.} 
  \label{fig:wall}   
\end{figure}

Given the mass, the 1-particle form factor can be obtained from the 
wall-wall correlator by fitting the prefactor in (\ref{eq:G0-2p}). 
The result at $\beta=0.613$ is shown in fig.~\ref{fig:wall}.  The 
long-distance expansion and finite-size effects describe the data very 
well.  The resulting 1-particle form factors are given in table\  
\ref{tab:sim}.  The error estimates were again obtained from a 
jackknife analysis which also included the determination of the mass 
described above.

\begin{figure}[htbp]   
  \centering   
  \includegraphics[width=1\epswid]{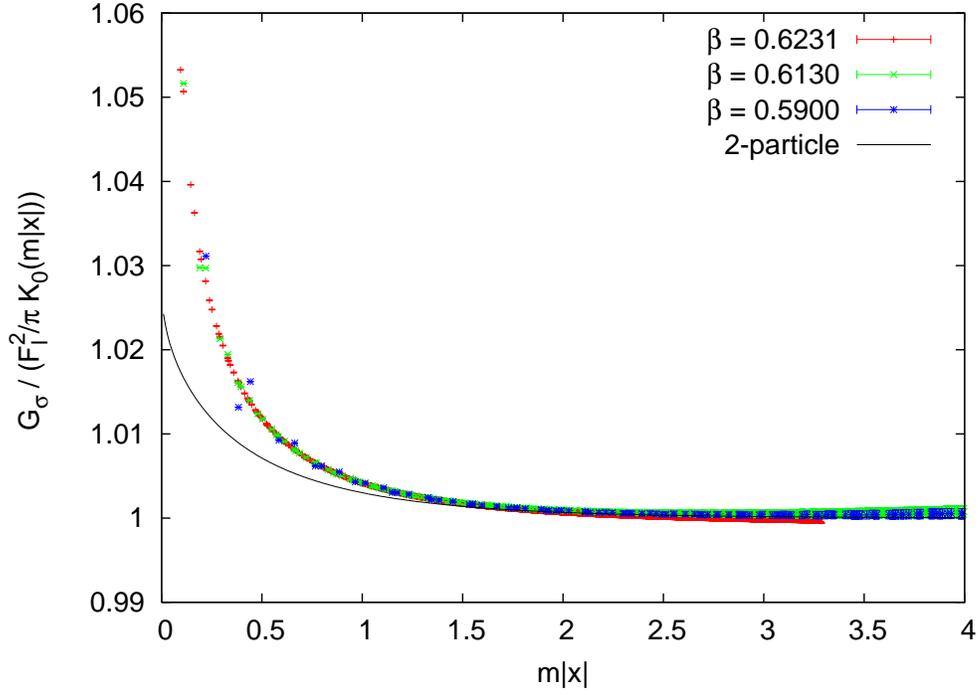}   
  \caption{Two-point function, rescaled using $m$ and  
    $F_{\ell}$ from fits to the wall-wall correlator and 
    divided by 1-particle term of long-distance expansion.} 
  \label{fig:cf}    
\end{figure}

The long-distance expansion of the 2-point function is completely 
determined by $m$ and $F_{\ell}$.  Figure\ \ref{fig:cf} shows 
the 2-point function divided by the 1-particle term of the 
long-distance expansion and rescaled by $m$ and $F_{\ell}$. 
The data from the three different lattices collaps nicely, and the 
small deviations from the 1-particle term are well described by the 
2-particle term. 
 
\subsection{Scaling violations}    
\label{sect4.2}   
   
Comparing the results from the different lattices, we may estimate the 
size of the scaling violations which we must expect in our results. 
 
\begin{figure}[tbhp] 
  \centering 
  \includegraphics[width=\epswid]{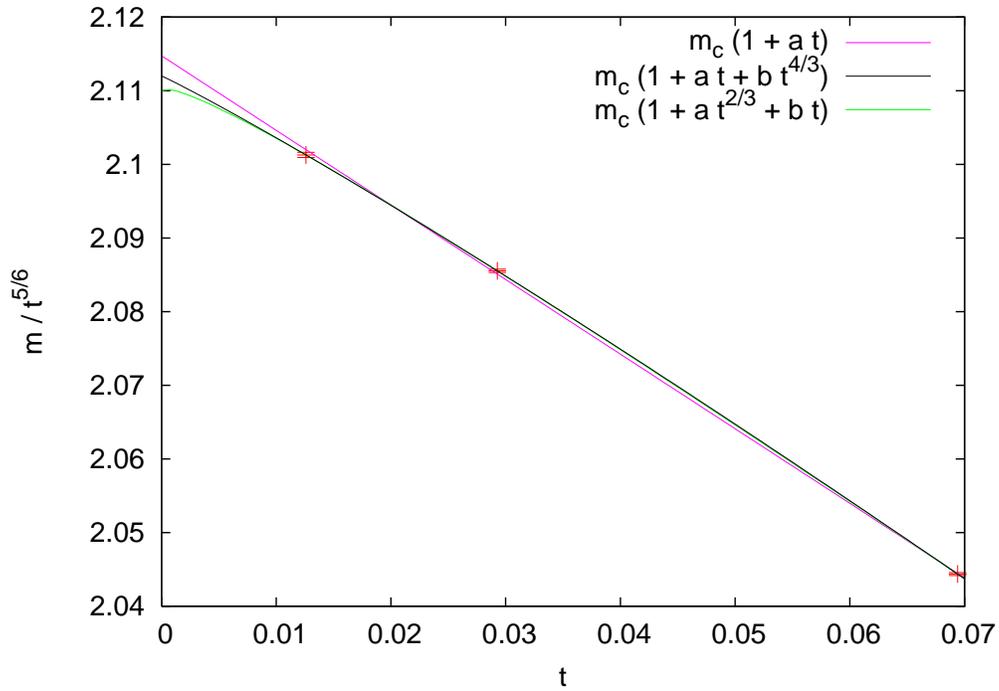} 
  \caption{Correction to scaling for $m$.} 
  \label{fig:m} 
\end{figure}

\begin{figure}[ht] 
  \centering 
  \includegraphics[width=\epswid]{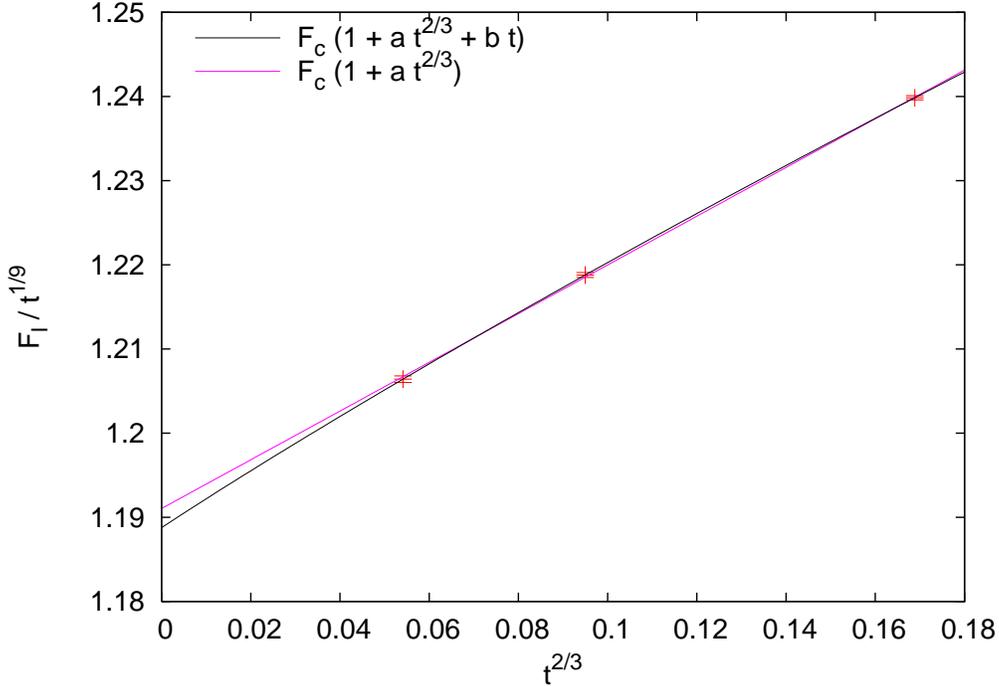} 
  \caption{Correction to scaling for $F_{\ell}$.} 
  \label{fig:F} 
\end{figure}

Setting   
\begin{equation} 
t\equiv \frac{T-T_c}{T_c} = \frac{\beta_c-\beta}{\beta}   
\end{equation} 
we expect $m\sim t^{5/6}$ and $F_\ell\sim t^{1/9}$.  Figures 
\ref{fig:m} and \ref{fig:F} show the ratios $m/t^{5/6}$ and 
$F_\ell/t^{1/9}$ as a function of $t$.  They vary by about $3\%$, a 
reasonable size for scaling violations at these correlation lengths. 
 
The scaling violations in the mass are expected to be of the form 
\begin{align} 
  m(t)/t^{5/6} &= m_{c} (1 + a_{m,\Delta} t^{\Delta} + a_{m,1} t + \dots)\,.  
\label{eq:scalingviolation} 
\end{align} 
Taking into account that the subleading thermal operator $\epsilon'$ corresponds to the conformal operator  
$(3,1;3,1)$, the leading correction-to-scaling exponent is $\Delta=-\frac{2-X_{\epsilon'}}{2-X_\epsilon}=\frac23$.  
This value has been first determined in \cite{Nienhuis} and later confirmed in 
\cite{AdlerPrivman,vonGehlen,deQueiroz}, see \cite{Shchur}.  We find 
$a_{m,\Delta}\ll a_{m,1}$.  A fit with only one term and variable 
exponent $p$ yields $p=1.10\dots$.  These two facts indicate that 
$a_{m,\Delta}$ may vanish.  A fit with only the linear term has 
$\chi^2=7.7$, so we perform a fit with the next two terms, 
\begin{align} 
  m_{c} (1 + a_{m,1} t + a_{m,2\Delta} t^{2\Delta}) 
\;. 
\end{align} 
Since it has as many parameters as data points, we use the fit with 
only the linear term for an error estimate.  The result is 
\begin{equation} 
  m_{c}=2.112(3) 
\;. 
\label{eq:mc} 
\end{equation} 
The statistical error from the fit with only a linear term is much 
smaller.  The fits are shown in fig.~\ref{fig:m}. 
 
For $F_{\ell}$, a fit to 
\begin{align} 
  F_\ell(t)/t^{1/9} &= F_{c} (1 + a_{F,\Delta} t^{\Delta} + a_{F,1} t + \dots)  
\label{eq:scalingviolationF} 
\end{align} 
yields a linear term which is much smaller than the leading 
correction.  We use a fit with only the leading correction for an 
error estimate and obtain 
\begin{equation} 
  F_{c}=1.189(3) 
\;. 
\label{eq:Fc} 
\end{equation} 
Again the statistical errors from the latter fit are smaller than the 
quoted error.  Both fits are shown in fig.~\ref{fig:F}.

\subsection{3-point function}   
\label{sect4.4}   
\begin{figure}[htbp]   
  \centering   
  \includegraphics[width=1\epswid]{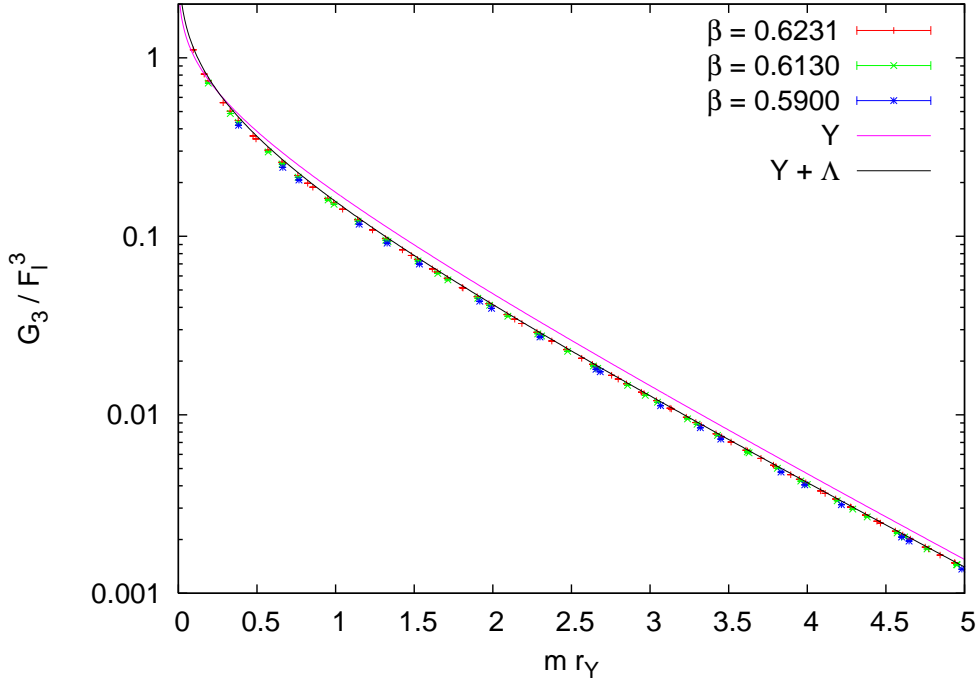}   
  \caption{Three-point function, rescaled using $m$ and  
    $F_{\ell}$ from fits to the wall-wall correlator.} 
  \label{fig:3-point}   
\end{figure}

The long-distance expansion of the 3-point function is completely 
determined by $m$ and $F_{\ell}$.  In particular, 
$G_{\ell}^{(3)}/F_{\ell}^3$ as a function of $m x_i$ is independent 
of any parameters.  The measured 3-point functions, rescaled in this 
way, are shown in fig.\ \ref{fig:3-point}.  Also shown is the 
long-distance expansion of eq.\ (\ref{g3equi}) which we split into 
``Y-type'' and ``$\Lambda$-type'' contributions (the second term 
depends on sums of two spin--spin-distances), 
\begin{equation}   
  \label{eq:G3}   
  G_{\ell}^{(3)}(x_1, x_2, x_3) = F_{\ell}^3 \Gamma^{\bar A}_{A A} \bigl[ 
  g_{\textrm{Y}}(x_1, x_2, x_3) + g_{\Lambda}(x_1, x_2, x_3)  \bigr] 
\end{equation}   
with   
\begin{align}   
  g_{\textrm{Y}}(x_1, x_2, x_3) &= \frac{1}{\pi} K_0(m r_Y)   
  \\   
  g_{\Lambda}(x_1, x_2, x_3) &= \frac{3}{2\pi^2}   
  \int_{-\infty}^{\infty} \dd\theta\, \frac {F^\sigma_{A   
      A}(\theta+\ii\pi/3)} {F^\sigma_{\bar A} \Gamma^{\bar A}_{A A}}   
  K_0\left( 2 m R \cosh\frac{\theta}{2} \right)   
\end{align}   
where $R=|x_1-x_2|=|x_2-x_3|=|x_3-x_1|$ is the distance between two 
spins and $r_Y=\sqrt3 R$ the Y-length.  For the equilateral geometry, 
the Y-type term is leading.  It is shown separately in the figure. 
The subleading $\Lambda$-type term is sizable up to large distances. 
This is not unexpected as the sum of two sides, $2R$, is only 15\% 
larger than the Y length $\sqrt{3}R$. 
   
\begin{figure}[htbp]   
  \centering   
  \includegraphics[width=1\epswid]{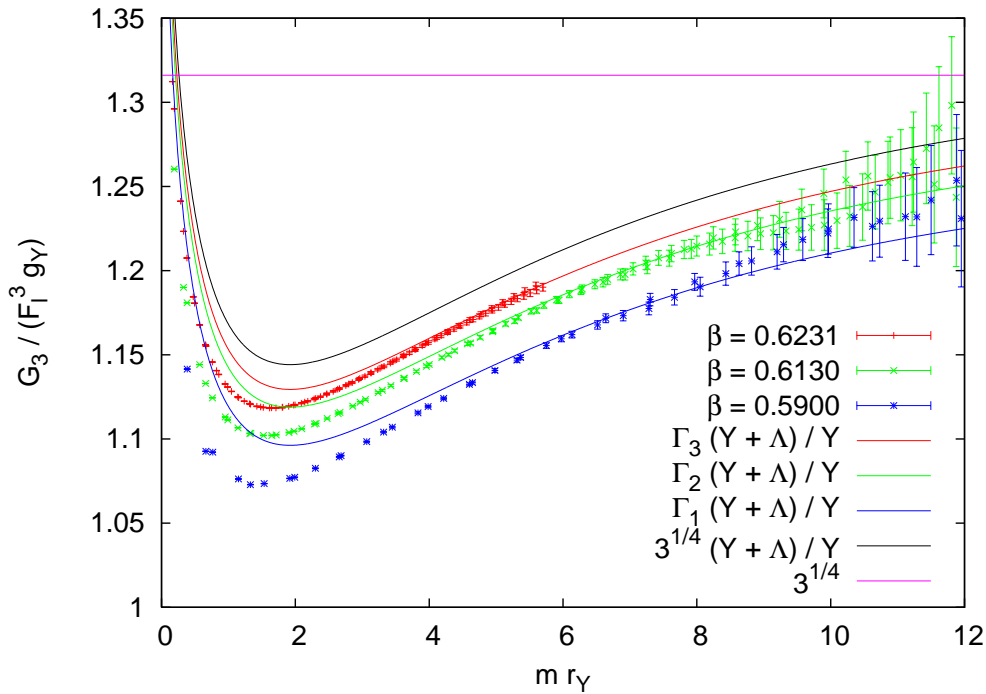}      
  \caption{Same as fig.~\ref{fig:3-point}, but divided by the leading 
    (Y-type) contribution.  Note the larger range.}   
  \label{fig:3-point-divided}   
\end{figure}

The deviation from the leading Y-type term is shown in 
fig.~\ref{fig:3-point-divided}.  Scaling violations are clearly visible 
and turn out to be of the same order as those observed in the 2-point 
function, i.e.\ about $3\%$.  These scaling violations have to be 
expected (even after rescaling with $m$ and $F_{\ell}$) since the 
lattice model at a given temperature $t$ is effectively described by a 
quantum field theory which differs from the scaling theory 
(\ref{Stau}) by irrelevant operators.  The form factors therefore 
differ from the values computed exactly in the integrable scaling 
theory.  The two terms above contain the three-particle coupling 
constant $\Gamma^{\bar A}_{A A}$ and the 2-particle form factor 
$F^{\sigma}_{A A}(\theta)$ which is also proportional to $\Gamma^{\bar 
  A}_{A A}$.  Lacking an understanding of the rapidity-dependence of 
the scaling violations, we proceed by ignoring all scale-dependence in 
$F^\sigma_{A A}$ other than that induced by $\Gamma^{\bar A}_{A A}$, 
i.e.\ we replace the prefactor $\Gamma^{\bar A}_{A A}$ in 
(\ref{eq:G3}) by a function of $t$.  This works very well, as can be 
seen from the figure.  The remaining deviations may well be due to 
higher-order terms in the long-distance expansion. 
The resulting values for the three-particle coupling are given in 
table~\ref{tab:sim}.  The quoted errors were taken from the 3-point 
function at distances where $G_{\ell}^{(3)}/(g_Y+g_\Lambda)$ 
approaches a constant.  They are therefore only rough estimates.  The 
values of $\Gamma^{\bar A}_{A A}(t)$ nicely approach the value 
$3^{1/4}=1.316\dots$ in the scaling theory.  The corresponding curve 
is also shown in the figure.  Note that this curve does not depend on 
any fits, but is completely determined by the values of $m$ and 
$F_\ell$ extracted from the wall-wall correlator. 
   
\section{Short distance expansion}   
\label{sect5}   
\subsection{2-point function}    
In order to have good control on the $\langle \sigma \bar \sigma \rangle$ correlation function   
it is worth to compare the Monte-Carlo data in the short distance   
regime with the corresponding perturbative expansion obtained in the   
framework of the so-called Conformal Perturbation Theory.    
Since such a perturbative expansion is expected to be valid in the region   
$m |x| \ll 1$, it is a complementary tool with respect to the form factor    
expansion discussed before.   
    
Let us recall the main results about Conformal Perturbation Theory in the special case    
of the correlation function $\langle \sigma(x) {\bar \sigma}(0) \rangle$.    
Following the standard literature on the subject    
\cite{Zamolodchikov:1990bk}-\cite{Guida:1996nm}    
we may write such a correlator as   
\bea   
G_\sigma(x) =  \langle \sigma(x) {\bar \sigma}(0) \rangle =    
\sum_p {\mathcal C}_{\sigma {\bar \sigma}}^{[\phi_p]} (x; \tau)    
\langle [\phi_p] \rangle   
\eea   
here the sum over $p$ ranges over all the conformal families allowed    
by the Operator Product Expansion of $\sigma$ and $\bar \sigma$.      
The Wilson coefficients ${\mathcal C}_{\sigma {\bar \sigma}}^{[\phi_p]}$    
can be calculated perturbatively in the coupling constant $\tau$.   
Their Taylor expansion (w.r.t. $\tau$) is:   
\bea   
 {\mathcal C}_{\sigma {\bar \sigma}}^{[\phi_p]}(x;\tau) = \sum_k    
 \frac{\tau^k}{k!} \partial^k_\tau    
 {\mathcal C}_{\sigma {\bar \sigma}}^{[\phi_p]}(x;0).   
\eea     
It is possible to show that the   
derivatives of the Wilson coefficients which appear in the previous expression    
can be written in terms of multiple integrals of the conformal correlators   
\bea   
\partial^k_\tau    
 {\mathcal C}_{\sigma {\bar \sigma}}^{[\phi_p]}(x;0) = 
(-1)^k \int^{'} d^2 z_1 \dots d^2 z_k \,   
 \langle \sigma(x) {\bar \sigma(0)} \epsilon(z_1) \dots \epsilon(z_k) [\phi_p](\infty)    
 \rangle_{\textrm{\tiny cft}}   
\eea    
where the prime on the integral implies a suitable treatment of the IR divergences and we used a shorthand    
notation to denote the follwoing limit:   
\bea   
\langle \sigma(x_1) .....   
\epsilon(\infty)  \rangle_{\textrm{\tiny cft}}=    
\lim_{w \to \infty} \frac{\langle \sigma(x_1) .....    
\epsilon(w)  \rangle_{\textrm{\tiny cft}}}{\langle \epsilon(w)  \epsilon(0)   
  \rangle_{\textrm{\tiny cft}}}~~~.   
\eea   
For all the details we address the interested reader to the original literature    
\cite{Guida:1995kc}.   
   
Another important piece of information which enters the perturbative expansion of   
the correlator is represented by the Vacuum Expectation Values $\langle [\phi_p] \rangle$.   
Such VEVs cannot be calculated in the framework of the perturbation theory,   
and being non-perturbative objects they have to be obtained by other methods.   
By using the powerful techniques of Integrable QFTs, they were computed    
in a series of papers \cite{Lukyanov:1996jj}-\cite{Fateev:1998xb} for a wide    
class of theories, including various   
integrable perturbations of the Minimal Models.   
   
The leading term of the perturbative expansion is given by the 2-point conformal   
correlator which corresponds to the choice $k=0$ and $\phi_p= \mathbb I$    
\bea   
{\mathcal C}_{\sigma {\bar \sigma}}^{ \mathbb I}(x;0)= \frac{1}{|x|^{4/15}}   
\label{2pt} 
\eea   
which follows from the choice of the conformal normalization    
$C_{\sigma {\bar \sigma}}^{ \mathbb I}=1$.   
   
The first few sub-leading terms are    
\bea   
G_\sigma(x) =  \langle \sigma(x) {\bar \sigma}(0) \rangle =    
{\mathcal C}_{\sigma {\bar \sigma}}^{ \mathbb I}(x;\tau) +   
{\mathcal C}_{\sigma {\bar \sigma}}^{ \epsilon}(x;\tau)    
\langle \epsilon \rangle+ \dots   
\eea   
where   
\bea   
{\mathcal C}_{\sigma {\bar \sigma}}^{ \mathbb I}(x;\tau) & = &    
{\mathcal C}_{\sigma {\bar \sigma}}^{ \mathbb I}(x;0) + \tau \,   
\partial_\tau {\mathcal C}_{\sigma {\bar \sigma}}^{ \mathbb I}(x;0) + \dots   
\nonumber \\   
{\mathcal C}_{\sigma {\bar \sigma}}^{ \epsilon}(x;\tau) & = &   
{\mathcal C}_{\sigma {\bar \sigma}}^{ \epsilon }(x;0) + \dots,   
\eea   
give corrections up to $\tau$. The explicit   
expression of the various contributions is, together with (\ref{2pt}),  
\bea    
\partial_\tau {\mathcal C}_{\sigma {\bar \sigma}}^{ \mathbb I}(x;0) & = &   
- \int d^2 z  \langle \sigma(x) {\bar \sigma(0)} \epsilon(z) \rangle_{\textrm{\tiny cft}}   
=-C_{\sigma {\bar \sigma}}^\epsilon \, |x|^{14/15} \int d^2 y |y|^{-4/5} |1-y|^{-4/5}   
\nonumber \\   
&=& \sin \frac{4 \pi}{5} \left| \frac{\Gamma(-1/5)\Gamma(3/5)}{\Gamma(2/5)} 
\right|^2   
C_{\sigma {\bar \sigma}}^\epsilon \, |x|^{14/15}    
\nonumber \\   
{\mathcal C}_{\sigma {\bar \sigma}}^{ \epsilon }(x;0) &= &C_{\sigma {\bar \sigma}}^\epsilon  \,|x|^{8/15}   
\eea   
where the Wilson coefficient    
\bea   
C_{\sigma {\bar \sigma}}^\epsilon &=&\frac12 \sqrt{\frac{\sin \frac{2 \pi}{5} }{\sin  \frac{ \pi}{5}} }\,  
 \frac{\Gamma^2(3/5)}{\Gamma(2/5)\Gamma(4/5)} = 0.546178 \dots   
\eea   
can be found combining the results of \cite{Dotsenko:1985hi} and \cite{Klassen:1991dz}   
(see also \cite{McCabe:1995cf}).       
The integral which appears in    
$\partial_\tau {\mathcal C}_{\sigma {\bar \sigma}}^{ \mathbb I}(x;0) $ is    
well known. It is a particular case of the following one   
\bea   
{\mathcal Y}_{a,b} = \int d^2 z |z|^{2a} |1-z|^{2b} = \frac{\sin \pi (a+b) \, \sin \pi b}{\sin \pi a}   
 \left| \frac{\Gamma(-a-b-1)\Gamma(b+1)}{\Gamma(-a)}   
\right|^2~~~~.   
\eea   
Its value is: ${\mathcal Y}_{-\frac25,-\frac25} = -8.97743 \dots$.    
   
At last, let us discuss the other (non-perturbative) quantities we need:    
the VEV of the perturbing operator   
$\epsilon(x)$ and the relation between the coupling constant and the mass   
of the fundamental particle. The latter is known, and can be extracted from    
\cite{Fateev:1993av}   
\bea   
\tau = \kappa\, m^{6/5}, \ \ \ \ \kappa = 0.164303 \dots.   
\label{kappa} 
\eea   
The VEV $\langle \epsilon \rangle$ can be easily computed starting from the   
knowledge of the vacuum energy density \cite{Zamolodchikov:1989cf}   
\bea   
\varepsilon_0 = - \frac{\sqrt 3}{6} m^2    
\eea   
which is related to the VEV of the perturbing operator by means of the    
relation   
\bea    
\langle \epsilon \rangle =  \partial_\tau \varepsilon_0 =   
A_\epsilon\, \tau^{2/3} =- 9.761465 \dots \tau^{2/3} =- 2.92827\dots m^{4/5}.   
\eea   
Collecting the above ingredients, the perturbative series can be cast in the    
following form   
\bea   
G_\sigma(x) & = &    
\frac{1}{|x|^{4/15}} \left(1 +    
C_{\sigma {\bar \sigma}}^\epsilon  A_\epsilon \, \tau^{2/3} |x|^{4/5} +   
{\mathcal Y}_{-\frac25,-\frac25} C_{\sigma {\bar \sigma}}^\epsilon \, \tau |x|^{6/5}    
+\dots \right)    
\nonumber \\   
& =  & \frac{1}{|x|^{4/15}} \left(1 +    
C_{\sigma {\bar \sigma}}^\epsilon  A_\epsilon \, u^{2/3}  +   
{\mathcal Y}_{-\frac25,-\frac25} C_{\sigma {\bar \sigma}}^\epsilon \, u   
+\dots \right)    
\nonumber \\   
& =  &\frac{1}{|x|^{4/15}} \left(1 +    
C_{\sigma {\bar \sigma}}^\epsilon  A_\epsilon  \kappa^{2/3} \, r^{4/5} +   
{\mathcal Y}_{-\frac25,-\frac25} C_{\sigma {\bar \sigma}}^\epsilon \,    
\kappa \, r^{6/5} +\dots \right)    
\nonumber   
\eea    
where we set $u=\tau |x|^{6/5}$, and $r = m |x|$, respectively.   
   
\subsection{3-point function}     
The perturbative approach of the previous section can be generalized    
to multipoint correlators as well. Following the guidelines of \cite{Guida:1996nm}   
one is able to write down a perturbative IR safe expansion  
\bea   
G^{(n)} (x) =  \langle \phi_1(x_1) \dots \phi_n(x_n) \rangle =    
\sum_p {\mathcal C}_{ \phi_1 \dots \phi_n}^{[\phi_p]}    
(\underline{x}; \tau)    
\langle [\phi_p] \rangle   
\eea   
where the structure functions ${\mathcal C}_{ \phi_1 \dots \phi_n}^{[\phi_p]} (\underline{x}; \tau) $ are a generalization of those appearing in the formula for the   
two-point function\footnote{We used the notation    
$\underline{x} = \{x_1, \dots, x_n\}$.}. Their Taylor expansion gives   
\bea   
{\mathcal C}_{ \phi_1 \dots \phi_n}^{[\phi_p]} (\underline{x}; \tau) =    
{\mathcal C}_{ \phi_1 \dots \phi_n}^{[\phi_p]} (\underline{x}; 0) +   
\sum_{\ell =1}^{\infty} \frac{\tau^\ell}{\ell !}    
\partial_\tau^\ell {\mathcal C}_{ \phi_1 \dots \phi_n}^{[\phi_p]} (\underline{x}; 0)    
\eea   
where we have   
\bea   
{\mathcal C}_{ \phi_1 \dots \phi_n}^{[\phi_p]} (\underline{x}; 0) =    
\langle \phi_1(x_1) \dots \phi_n(x_n) [\phi_p](\infty)  \rangle_{\textrm{\tiny cft}},   
\eea   
and   
\bea   
\partial_\tau^\ell {\mathcal C}_{ \phi_1 \dots \phi_n}^{[\phi_p]} (\underline{x}; 0)=(-1)^{\ell} 
\int^{'} d^2 z_1 \dots d^2 z_\ell \,    
\langle \phi_1(x_1) \dots \phi_n(x_n) \epsilon(z_1) \dots \epsilon(z_\ell)   
 [\phi_p](\infty)  \rangle_{\textrm{\tiny cft}}   
\eea   
where the operator $\epsilon(x)$ which appears in the previous expression is the perturbing   
operator conjugated to the coupling constant $\tau$.   
   
In the present section we are interested in the 3-point function   
$G^{(3)} = \langle \sigma_1 \sigma_2 \sigma_3 \rangle$, whose perturbative expression 
can be written according to the previous considerations as    
\bea   
G^{(3)} (x) &  = & \langle \sigma(x_1) \sigma(x_2)  \sigma(x_3) \rangle =   
\nonumber \\   
& = &    
{\mathcal C}_{ \sigma \sigma \sigma}^{{\mathbb I}}    
(x_1,x_2,x_3; \tau) +{\mathcal C}_{ \sigma \sigma \sigma}^{\epsilon}    
(x_1,x_2,x_3; \tau) \, \langle \epsilon \rangle+    
\dots  \,\,; 
\eea    
up to first order in $\tau$ one has   
\bea   
{\mathcal C}_{ \sigma \sigma \sigma}^{{\mathbb I}}    
(x_1,x_2,x_3; \tau) & = &   
{\mathcal C}_{ \sigma \sigma \sigma}^{{\mathbb I}}    
(x_1,x_2,x_3; 0) +  \tau\, \partial_\tau {\mathcal C}_{ \sigma \sigma \sigma}^{{\mathbb I}}    
(x_1,x_2,x_3; 0) + \dots    
\nonumber \\   
{\mathcal C}_{ \sigma \sigma \sigma}^{\epsilon}    
(x_1,x_2,x_3; \tau) & = &   
{\mathcal C}_{ \sigma \sigma \sigma}^{\epsilon}    
(x_1,x_2,x_3; 0) +\tau\,  \partial_\tau   
{\mathcal C}_{ \sigma \sigma \sigma}^{\epsilon}     
(x_1,x_2,x_3; 0) + \dots   
\nonumber   
\eea   
The explicit expression of the zero-th order contributions can be derived quite   
easily. We have   
\bea   
{\mathcal C}_{ \sigma \sigma \sigma}^{{\mathbb I}}    
(x_1,x_2,x_3; 0) &=& \langle \sigma(x_1) \sigma(x_2)  \sigma(x_3)  \rangle_{\textrm{\tiny cft}} =    
\nonumber \\   
&=& \frac{C^{\bar \sigma} _{\sigma \sigma}}{|x_1-x_2|^{2/15}|x_2 -x_3|^{2/15}   
|x_1-x_3|^{2/15}}   
\label{3pt} 
\eea   
where the structure constant $C^{\bar \sigma} _{\sigma \sigma}$ can be found in the   
literature \cite{Dotsenko:1985hi,Klassen:1991dz,McCabe:1995cf}   
\bea   
C^{\bar \sigma} _{\sigma \sigma}=    
\sqrt{\frac{3 \sin2 \pi/5 }{\pi \sin \pi/5}}    
\frac{\Gamma(5/6)\Gamma^2(3/5)\Gamma^4(1/3)}{\Gamma(2/5)\Gamma(4/5)\Gamma^2(2/3)\Gamma^2(1/6)}   
=1.09236 \dots.   
\eea   
The other zero-th order term is given by   
\bea   
{\mathcal C}_{ \sigma \sigma \sigma}^{\epsilon}    
(x_1,x_2,x_3; 0)  =    
\langle \sigma(x_1)  \sigma(x_2) \sigma(x_3) \epsilon(\infty)  \rangle_{\textrm{\tiny cft}}   
\eea   
whose main ingredient is the conformal four point correlator   
$\langle \sigma\sigma\sigma\epsilon \rangle$ which was computed in    
\cite{Gliozzi:1997yc}. We consistently fixed its normalization constant,   
which gives    
\bea   
\langle \sigma(x_1)  \sigma(x_2) \sigma(x_3) \epsilon(x_4)  \rangle_{\textrm{\tiny cft}}   
=C^{\bar \sigma} _{\sigma \sigma} C^{\epsilon} _{\sigma {\bar \sigma}} \,   
\frac{|x_{12}x_{13}x_{23}|^{2/15}}{|x_{14}x_{24}x_{34}|^{8/15}} \,   
|y(1-y)|^{14/15} \,   
\left \{   
|f_1(y)|^2 + K^{-1} |f_2(y)|^2   
\right \}   
\eea   
where   
\bea   
f_1(y) = y^{-3/5} {_2F_1}(1/5,4/5;2/5;y), \ \ \ f_2(y) = {_2F_1}(4/5,7/5;8/5;y)   
\eea   
and   
\bea   
K = \frac{9\, \Gamma(1/5) \, \Gamma(3/5)^3 }{4\, \Gamma(4/5) \, \Gamma(2/5)^3 }, \ \ \ y=\frac{x_{14}x_{23}}{x_{34} x_{21}}, \ \ \ 1-y=\frac{x_{13}x_{24}}{x_{34} x_{12}}   
\eea   
Then, a simple computation gives   
\bea   
{\mathcal C}_{ \sigma \sigma \sigma}^{\epsilon}    
(x_1,x_2,x_3; 0)  =    
C^{\bar \sigma} _{\sigma \sigma} C^{\epsilon} _{\sigma {\bar \sigma}} \,
|x_{12}x_{13}x_{23}|^{2/15} |{\tilde y}(1-{\tilde y})|^{14/15} \,   
\left \{   
|f_1({\tilde y})|^2 + K^{-1} |f_2({\tilde y})|^2   
\right \}   
\eea   
where   
\bea   
{\tilde y}=\frac{x_{23}}{x_{21}}, \ \ \ 1-{\tilde y}=\frac{x_{13}}{x_{12}}   
\eea   
The first order terms are as follows    
\bea   
\partial_\tau {\mathcal C}_{ \sigma \sigma \sigma}^{{\mathbb I}}    
(x_1,x_2,x_3; 0) & = & - \int^{'} d^2 z    
\langle \sigma(x_1)  \sigma(x_2) \sigma(x_3) \epsilon(z)  \rangle_{\textrm{\tiny cft}}   
\nonumber \\   
\partial_\tau {\mathcal C}_{ \sigma \sigma \sigma}^{{\epsilon}}    
(x_1,x_2,x_3; 0) & = &  - \int^{'} d^2 z   
\langle \sigma(x_1)  \sigma(x_2) \sigma(x_3) \epsilon(w)  \epsilon(\infty)  \rangle_{\textrm{\tiny cft}}   
\eea   
but we did not manage to calculate them analytically for a generic triangle.   
We expect that some simplifications will occur when a given geometry is chosen,    
for example in the case of equilateral triangles.

\section{Comparison with Monte-Carlo data at short distance}   
\label{sect6}   
\subsection{2-point function}   
\label{sect6.1}   
The short distance expansion calculated in section \ref{sect5} can be    
compared with the Monte-Carlo data in the region of short distances,  
i.e.\ $m |x| \ll 1$.   
   
For this purpose, it is first necessary to rewrite such an expansion in the    
dimensionless variable $r=m |x|$   
\bea   
{\tilde G_\sigma (r)} = m^{-4/15} \langle \sigma(x) {\bar \sigma}(0) \rangle =   
\frac{1}{r^{4/15}} \left(1+g_1 r^{4/5}+g_2 r^{6/5}+g_3 r^{2}+g_4 r^{12/5}+ \dots \right)   
\label{shdis}   
\eea   
where, from the previous section we have    
\bea   
g_1=C_{\sigma {\bar \sigma}}^\epsilon  A_\epsilon  \kappa^{2/3}=-1.59936 \dots, \ \ \ \ g_2={\mathcal Y}_{-\frac25,-\frac25} C_{\sigma {\bar \sigma}}^\epsilon    
\kappa =0.805622 \dots,    
\eea   
and $g_3$ and $g_4$ are unknown constants which embody the contributions   
given by higher orders in perturbation theory.   
   
We chose to use the sample of Monte-Carlo data with the largest correlation length,   
i.e.\ $\xi = 18.232(7)$. This is motivated by the following requirements:   
\begin{itemize}   
\item{The perturbative calculation is supposed to hold for $r \ll 1$;}   
\item{The region of very short distances ($|x|\lesssim5$, i.e.\ $r=|x|/\xi\lesssim0.27$)    
have to be avoided because    
of the so-called lattice artifacts, which are non-universal corrections induced by the   
lattice when the distance is comparable with the lattice spacing (see    
\cite{Caselle:2001zd} for an   
analysis of this point in the context of the Ising model in magnetic field).}   
\end{itemize}   
Such requirements imply the presence of a limited "window" in which the comparison   
between the data and the perturbative expansion can be done. Hence,   
the choice we made is motivated by the necessity of   
maximizing the number of data which fall in such a window in order to have    
a reliable sample for the fitting procedure.     

We used the following fitting function    
\bea   
{\tilde G_\sigma^\ell (r )} = m^{-4/15} \langle \sigma_\ell(x) {\bar \sigma}_\ell(0) \rangle =   
\frac{N_{\sigma}}{r^{4/15}} \left(1+g_1 r^{4/5}+g_2r^{6/5}+g_3 r^{2}+g_4 r^{12/5} \right).   
\label{fitshort} 
\eea   
where the constants $g_1$ and $g_2$ are known analytically and can be fixed    
exactly, while the constants $g_3$, $g_4$ which parametrize higher order contributions in the    
perturbative expansions are left as free parameters and are fixed by the fit.   
The overall constant $N_{\sigma}$ which takes into account  
the different normalization of $\sigma$ on the lattice and  
in the continuum (the normalization of the latter has been  
fixed in (\ref{shdis}) and is the usual conformal normalization) is also   
considered as a free parameter for the fitting procedure.    
 
It turns out that lattice effects are larger than the statistical 
errors up to distances $r\approx0.5$ once correlations between the 
data points are taken into account.  This makes it impossible to use 
the goodness of fit as a criterion for a successful description of the 
data and for error estimates.  Instead, we vary the fit range in order 
to estimate the effect of lattice artifacts, and include higher-order 
terms in order to estimate their effect.  For our best estimate of the 
parameters, we use the shortest distance ($r=0.3$) where lattice 
artifacts are smaller than (uncorrelated) statistical errors as a 
lower bound, and the largest distance where the fit still works well 
($r=0.8$) as an upper bound.  We then shift the range down (to  
$[0.14,0.5]$) or up (to $[0.4,1.0]$) to get a rough estimate of the 
uncertainties.  We also performed a fit with terms up to $g_6$.  We 
included two more orders because the exponents of the next two terms 
are close (16/5 and 18/5), so cancelations have to be expected.  The 
upper bound has been increased to the maximal value where the fit 
still works.  The resulting parameters are shown in table\  
\ref{tab:cffit}.  The largest uncertainties come from the lower bound, 
i.e.\ from lattice artifacts.  We therefore estimate the parameters as 
\begin{align} 
  N_\sigma &= 0.6576(4) \\ 
  g_3 &= 0.167(18) \\ 
  g_4 &= -0.130(23) 
\;. 
\end{align} 
It must be stressed that the values of the constants $g_3$, $g_4$ 
could be affected by systematic errors larger than the quoted errors 
due to possible cancellations between the two contributions.  The 
quoted uncertainties also do not include the effect of scaling 
violations as they were obtained from a single correlation length. 
The lattice artifacts on the coarser lattices are too strong to allow 
for an estimate of the scaling violations in $g_3$ and $g_4$.  It 
would be very interesting to fix this uncertainty by a direct 
calculation of these coefficient, which could be performed in the 
framework of Conformal Perturbation Theory but requires techniques 
more sophisticated than those discussed in this paper. 
 
\begin{figure}[ht] 
  \centering 
  \includegraphics[width=\epswid]{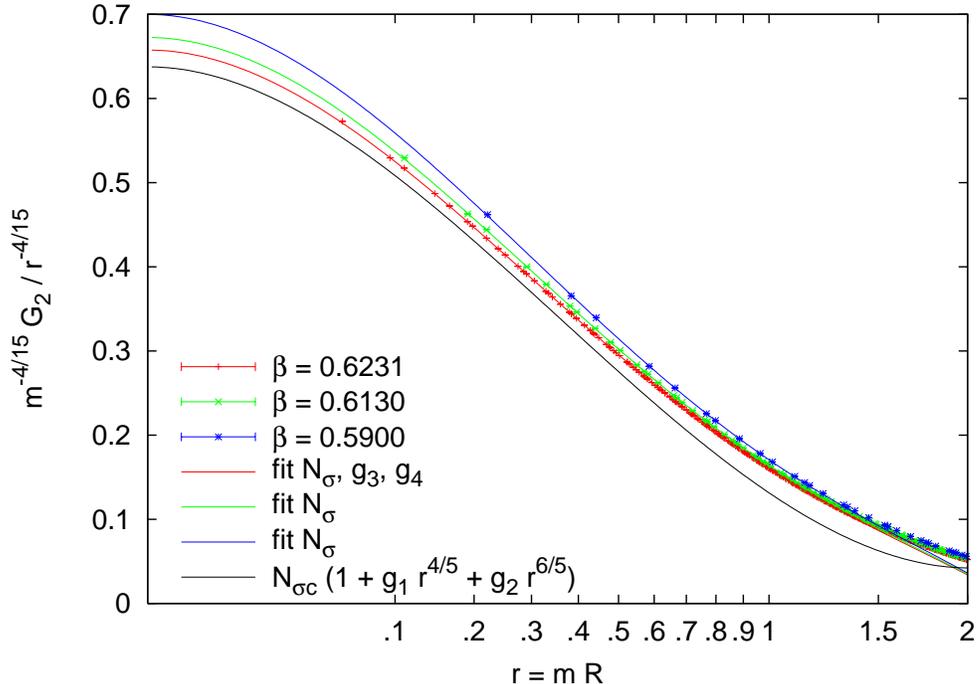} 
  \caption{2-point function divided by leading short-distance term  
    versus $r^{2/5}$.  Fit up to $g_4$ to $\beta=0.6231$ with range 
    $[0.3,0.8]$ (red) and series up to $g_4$ with only $N_\sigma$ 
    fitted to $\beta=0.613$ (green) and $\beta=0.59$ (blue) with 
    ranges $[0.4,0.8]$ and $[0.6,1.0]$, resp.  Theoretically known 
    part of series up to $g_2$ with prefactor from continuum 
    extrapolation of fig.\ \ref{fig:Ns} (black).} 
  \label{fig:cfshort} 
\end{figure}

We can get a rough estimate for the scaling violations in $N_\sigma$, 
by fitting it to the data on the coarser lattices while keeping $g_3$ 
and $g_4$ fixed.  Allowing for deviations due to the large lattice 
artifacts observed at small $r$, this works quite well.  The results 
are $N_\sigma=0.6725(8)$ for $\beta=0.613$ and $N_\sigma=0.6999(9)$ 
for $\beta=0.59$.  The uncertainties include \emph{only} the effect of 
varying the fit range.  We have no way of estimating the bias caused 
by keeping $g_3$ and $g_4$ fixed, let alone the effect of higher 
orders. 
The two fits as well as the above fit to the data at $\beta=0.6231$ 
are shown in fig.\ \ref{fig:cfshort}.  The rescaled correlation 
function $\tilde G_{\sigma}(r)$ has been divided by the leading  
short-distance term $r^{-4/15}$ for this figure.  Since the series has 
fractional exponents, we use $r^{2/5}$ for the abscissa. 
 
\begin{table} 
  \centering 
  \begin{tabular}{cccccccc} 
    parameters   & $x_{\text{min}}$ & $x_{\text{max}}$ & $N_\sigma$ & $g_3$ & $g_4$ & $g_5$ & $g_6$ \\ 
    \hline 
    $N_{\sigma},g_3,g_4$ & 0.3  & 0.8 & 0.6576 & 0.1674 & $-0.1297$ \\ 
    $N_{\sigma},g_3,g_4$ & 0.14 & 0.5 & 0.6580 & 0.1492 & $-0.1069$ \\ 
    $N_{\sigma},g_3,g_4$ & 0.4  & 1.0 & 0.6578 & 0.1641 & $-0.1261$ \\ 
    $N_{\sigma},g_3,g_4,g_5,g_6$ & 0.3  & 1.5 & 0.6574 & 0.1687 & $-0.1258$ & $-0.0386$ & 0.0242 \\ 
  \end{tabular} 
  \caption{Results of fits to 2-point function at $\beta=0.6231$.} 
  \label{tab:cffit} 
\end{table} 
 
\begin{figure}[ht] 
  \centering 
  \includegraphics[width=\epswid]{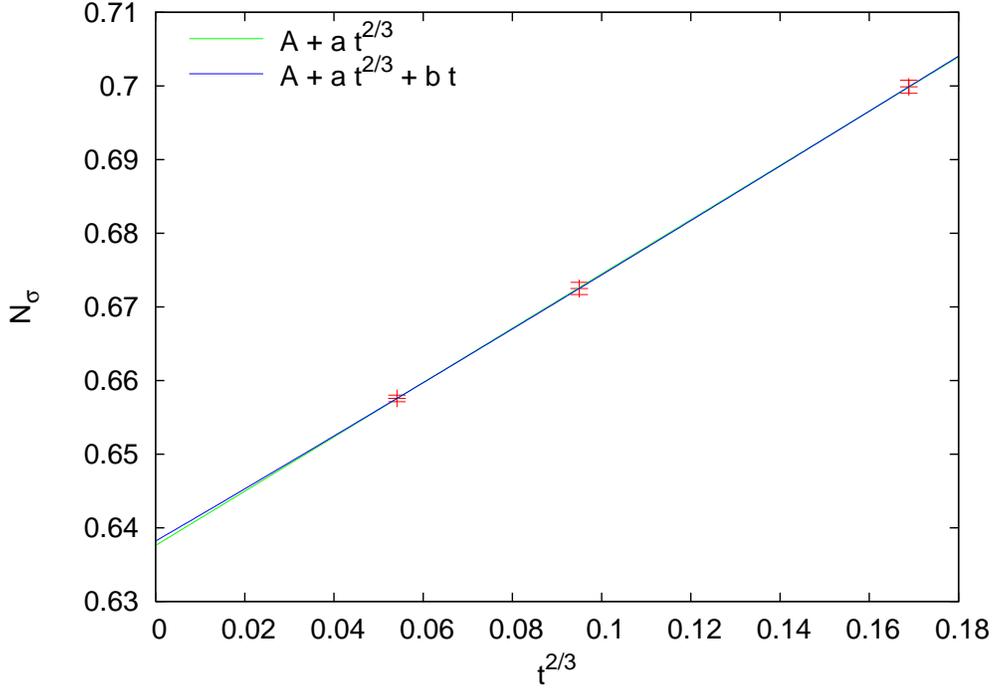} 
  \caption{Scale dependence of $N_\sigma$.} 
  \label{fig:Ns} 
\end{figure}

The scaling violations in $N_\sigma$ are expected to be of the form 
\begin{equation} 
  N_\sigma(t) = N_\sigma^c (1 + a_{N,\Delta} t^\Delta + \dots ) 
\end{equation} 
with the correction-to-scaling exponent $\Delta=2/3$ that already 
appeared in the discussion of the mass and the form factor in 
sect.~\ref{sect4.2}.  Remarkably, our crude estimates of $N_\sigma$ 
follow this prediction almost without deviation.  Figure~\ref{fig:Ns} 
shows a fit with $\Delta=2/3$ as well as a fit with two 
correction-to-scaling terms with exponents $2/3$ and $1$, which 
deviates only slightly.  From the former, we obtain a continuum 
extrapolation of 
\begin{align} 
  N_\sigma^c = 0.6376(6) 
\label{eq:Nc} 
\end{align} 
with the error taken as the difference of the two fits.  Because of 
the unknown systematic errors in the individual $N_\sigma$, this error 
is probably strongly underestimated. 
 
The short-distance expansion with this prefactor, and only including 
the analytically known terms up to $g_2$ is shown in 
fig.~\ref{fig:cfshort}.  The discrepancy between the data and this 
curve has two origins: scaling violations in the lattice data and 
omission of higher-order terms in the short-distance expansion.  In 
the continuum limit, and with higher-order terms included, the two are 
expected to meet.  The figure matches this expectation nicely. 
 
The constant $N_\sigma$ plays the role of the normalization    
of the magnetization operator $\sigma_\ell$ on the lattice,    
and for this reason it is not calculable in the framework of field theory.    
It is a non-universal quantity which depends,   
among other things, upon the geometry of the lattice.    
In the case of the Ising model on a square lattice    
the analogous quantity can be    
evaluated exactly thanks to the fact that the    
$\sigma \bar\sigma$ correlator 
can be evaluated analytically on the lattice at the critical point. By comparing   
this result with the continuum-limit expression the normalization constant is then easily obtained (see for   
instance \cite{Caselle:1999mg}). Unfortunately a similar exact lattice   
result does not exist for the three states Potts model on a triangular   
lattice.   
 
\subsection{Relation of short- and long-distance expansions} 
 
Given the normalization of the short-distance expansion $N_{\sigma}$, 
the amplitude of the 1-particle form factor $F^\sigma_{\bar A}$ can in 
principle be calculated in field theory using the cluster condition 
(\ref{cluster}).  The explicit form of $F^{\mu}_{\bar A A}(\theta)$ gives 
\begin{equation} 
  \label{eq:calN} 
  \mathcal{N} \equiv \frac{F^\sigma_{\bar A} F^{\bar\sigma}_A} {\langle\mu\rangle^2} 
  = 0.968240\dots 
\end{equation} 
The required amplitude of $\langle\mu\rangle$ in field theory, 
however, has not been computed so far.  We can give an estimate of the 
latter using the measured values of $N_{\sigma}$ and $F_{\ell}$. 
Since $\sigma_{\ell}\simeq\sqrt{N_{\sigma}}\sigma$, the lattice and field 
theory form factors are related as 
$F_{\ell}=\sqrt{N_{\sigma}}F^{\sigma}_{\bar A}$. 
 
In order to eliminate corrections to scaling, we extrapolate to the 
continuum limit.  The scale dependence on the lattice was given in 
sect.~\ref{sect4.2} in terms of $t$, 
\begin{align} 
  F_{\ell} &\sim F_{c} t^{1/9} 
\\ 
  m &\sim m_{c} t^{5/6} 
\end{align} 
while in field theory it is usually given in terms of the coefficient 
$\tau$ of the energy operator, 
\begin{align} 
  F^{\sigma}_{\bar A} &\sim A_F \tau^{1/9} 
\\ 
  m &\sim A_m \tau^{5/6} 
\;. 
\end{align} 
The ratio $F^{\sigma}_{\bar A}/m^{2/15}$ is scale-independent. 
The corresponding amplitude ratio can thus be identified on the 
lattice and in field theory, 
\begin{equation} 
  \frac{F_{c}/\sqrt{N_\sigma^c}} {m_{c}^{2/15}} 
= 
  \frac{A_F}{A_m^{2/15}} 
\;. 
\end{equation} 
 
The measured amplitudes (\ref{eq:mc}) and (\ref{eq:Fc}) together with 
(\ref{eq:Nc}) yield the ratio 
\begin{equation} 
  \frac{F_{c}/\sqrt{N_\sigma^c}} {m_{c}^{2/15}} = 1.348(4) 
\;. 
\end{equation} 
Using the known value $A_m=\kappa^{-5/6}=4.50431\dots$ from 
(\ref{kappa}), we get 
\begin{equation} 
  A_F = 1.647(5) 
\;. 
\end{equation} 
The amplitude of the disorder parameter is defined via 
\begin{equation} 
  \langle\mu\rangle \sim A_\mu \tau^{1/9} 
\;. 
\end{equation} 
The cluster condition (\ref{eq:calN}) then gives 
$A_\mu=A_F/\sqrt{\mathcal{N}}$ and thus 
\begin{equation} 
  A_\mu = 1.674(5) 
\;. 
\end{equation} 
All errors are probably underestimated, mainly because of the 
uncertainties in $N_\sigma^c$ as discussed in the previous section.

\subsection{3-point correlator: Equilateral triangles}   
Let us write explicitly the perturbative expansion of the 3-point    
correlator in the case of the equilateral triangle. It will be useful   
in the perspective of a comparison between Monte-Carlo data and the   
form factor expansion worked out previously.   
   
\begin{figure}[tbh] 
  \centering 
  \includegraphics[width=\epswid]{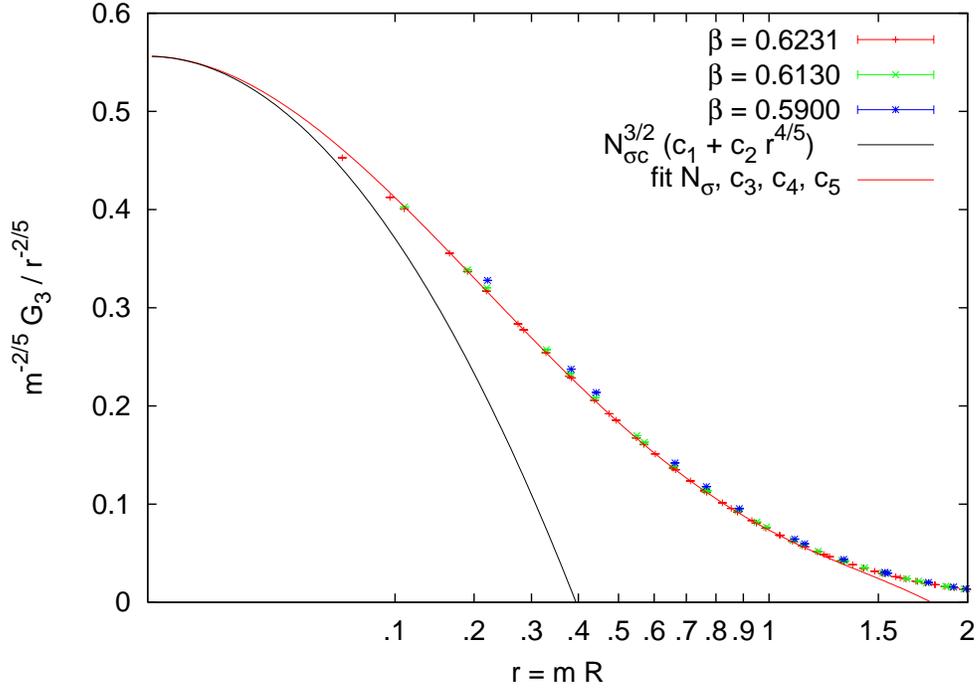} 
  \caption{3-point function divided by leading short-distance term,  
    theoretically known part of series up to $c_2$ with prefactor from 
    continuum extrapolation of $N_{\sigma}$ from 2-point function, and 
    fit up to $c_5$ to the data at $\beta=0.6231$ in the range 
    $[0.3,0.8]$.} 
  \label{fig:tpshort} 
\end{figure}

In such a particular case some simplifications occur, i.e.\ $|x_{i,j}| = R$   
where $R$ is the side of the triangle. Then, with the choice (obviously   
the final results will be independent of such a choice)   
\bea   
x_1=0, \ \ \ x_2 = R \, e^{i \pi /6}, \ \ \ x_3 = R \, e^{-i \pi /6}   
\eea   
we have    
\bea   
{\tilde y}=i \, e^{-i \pi /6} , \ \ \ 1-{\tilde y}=e^{-i \pi /3}   
\eea   
and finally   
\bea   
{\mathcal C}_{ \sigma \sigma \sigma}^{\epsilon}    
(0,R \, e^{i \pi /6},R \, e^{-i \pi /6}; 0)  =    
C^{\bar \sigma} _{\sigma \sigma} C^{\epsilon} _{\sigma {\bar \sigma}}    
 \,   
\left \{   
|f_1(i \, e^{-i \pi /6})|^2 + K^{-1} |f_2(i \, e^{-i \pi /6})|^2   
\right \}  R^{2/5} =  C^{\epsilon} _{\sigma \sigma \sigma } \, R^{2/5}    
\eea   
where   
\bea   
C^{\epsilon} _{\sigma \sigma \sigma } = 0.788825\dots.   
\eea   
Taking into account all the other pieces we have   
\bea   
G^{(3)}(R) =  \frac{1}{R^{2/5}} \left\{   
C^{\bar \sigma} _{\sigma \sigma}  +    
C^{\epsilon} _{\sigma \sigma \sigma } \, A_\epsilon \,   
\tau^{2/3} \, R^{4/5} + \dots   
\right\}.   
\eea   
Such an expression is very similar to the perturbative expansion of   
$\langle \sigma {\bar \sigma} \rangle$, in particular one can write it down using the   
dimensionless variable $u = \tau R^{6/5}$. As a consequence    
it is possible to guess the functional form of the higher order terms   
relying on dimensional analysis only. Hence we can write,    
in terms of $r = m R$,    
\bea   
G^{(3)}(R) =  \frac{1}{R^{2/5}} \left(   
c_1 + c_2 \, r^{4/5} + c_3 \, r^{6/5}  +c_4 \, r^2 + c_5 \, r^{12/5}  + \dots   
\right) 
\eea   
where 
\bea 
c_1 = C^{\bar\sigma} _{\sigma \sigma} = 1.09236\dots \;,\ \ \ \  
c_2 = C^{\epsilon} _{\sigma \sigma \sigma } \, A_\epsilon \, \kappa^{2/3} = -2.29795\dots 
\eea 
and $c_3,\dots$ are unknown (but in principle calculable) constants.  
It is also useful to define the scale invariant form of the correlator  
\bea   
\widetilde{G}^{(3)}(r) =m^{-2/5}\, \langle \sigma(x_1)  \sigma(x_2)  \sigma(x_3)  
\rangle=  
\frac{1}{r^{2/5}} \left( 
c_1 + c_2 \, r^{4/5} + c_3 \, r^{6/5}  +c_4 \, r^2 + c_5 \, r^{12/5}  + \dots   
\right)  
\eea     
and its corresponding expression on the lattice 
\bea   
\widetilde{G}^{(3)}_\ell(r) = 
\frac{N_{\sigma}^{3/2}}{r^{2/5}} \left( 
c_1 + c_2 \, r^{4/5} + c_3 \, r^{6/5}  +c_4 \, r^2 + c_5 \, r^{12/5}  + \dots   
\right) 
\eea 
where $N_{\sigma}$ is (the square of) the normalization of the lattice 
magnetization operator.  $N_{\sigma}$~has been extracted from the 
$\sigma\bar\sigma$ correlator in sect.~\ref{sect6.1}.  Since the 
scaling violations in the 3-point function are expected to be 
different from those in the 2-point function, we do not use the values 
of $N_{\sigma}$ fitted to the individual lattices, but rather the 
continuum extrapolation $N_\sigma^c$.  The series up to the 
known $c_2$ with this prefactor is shown in fig.\ \ref{fig:tpshort}. 
Note that the abscissa is the distance between two spins (in physical 
units and to the power $2/5$), not the Y-length $m 
r_{\textrm{Y}}=\sqrt{3}r$ which appears in the long-distance 
expansion.  The figure shows that, while higher-order terms are 
clearly important in the range where we have data, the leading term in 
the series is compatible with the latter. 
 
A reasonable description of the data on the finest lattice can be 
obtained with the series up to $c_5$ with $N_{\sigma}$ and 
$c_3,\dots,c_5$ as fit parameters.  Like in the case of the 2-point 
function, we use a fit range where both lattice artifacts are small 
compared to uncorrelated statistical errors ($r\ge0.3$) and the fit 
still works well ($r\le0.8$).  The fit is shown in the fig.\  
\ref{fig:tpshort}.  Contrary to the case of the 2-point function, the 
fit does not stay near the data beyond the fit range, and, not 
surprisingly, the parameters depend quite strongly on the latter, see 
table~\ref{tab:tpfit}.  Since the fit already includes three terms, we 
do not attempt to estimate the effect of higher-order terms by 
including even more terms.  The following uncertainties therefore 
contain neither the effects of these nor those of scaling violations: 
\begin{align} 
  N_\sigma &= 0.6376(36) \\ 
  c_3 &= 1.24(10) \\ 
  c_4 &= 0.44(45) \\ 
  c_5 &= -0.33(37) 
\;. 
\end{align} 
Still, it is quite remarkable that the fit yields a value of 
$N_{\sigma}$ almost identical to the one extracted from the 2-point 
function (the near coincidence of the curves at $r=0$ in the figure is 
not enforced!) 
 
\begin{table} 
  \centering 
  \begin{tabular}{cccccc} 
    $x_{\text{min}}$ & $x_{\text{max}}$ & $N_\sigma$ & $c_3$ & $c_4$ & $c_5$ \\ 
    \hline 
    0.3   & 0.8 & 0.6376 & 1.240 & $\phantom{-}0.441$ &           $-0.331$ \\ 
    0.15  & 0.5 & 0.6340 & 1.343 &           $-0.007$ & $\phantom{-}0.042$ \\ 
    0.4   & 1.0 & 0.6365 & 1.255 & $\phantom{-}0.393$ &           $-0.298$ 
  \end{tabular} 
  \caption{Results of fits to 3-point function at $\beta=0.6231$.} 
  \label{tab:tpfit} 
\end{table} 
 
Also contrary to the case of the 2-point function, the scaling 
violations on the coarser lattices cannot be described by just 
changing the prefactor $N_{\sigma}$.  We do not attempt fits with more 
parameters as there is not a window where both lattice artifacts are 
small and higher-order terms can be neglected.

\section{Implications for the three-quark potential}   
\label{sect7}   
 
In these last years much interest has been attracted by the study of the three-quark  
potential in Lattice Gauge Theories (LGT). Besides the obvious   
phenomenological interest of the problem, the three-quark potential is also a   
perfect tool for testing our understanding of the flux tube model of confinement   
 and of its theoretical description in terms of effective string models. These   
 models have been elaborated in the past years looking at the quark-antiquark   
 potential and their extension to baryonic states is highly non trivial.    
 Thanks to the improvement in lattice simulations (a summary of numerical   
 results can be found in~\cite{aft02}-\cite{mnst02}), the qualitative   
 behaviour of the three-quark potential is now rather well understood (for a   
 recent review see~\cite{fj05}).   
 \begin{itemize}   
 \item   
  For large interquark distances the three-quark   
 potential is well described by the so called {\bf Y} law which assumes    
 a flux tube configuration composed by    
 three strings which originate from the three quarks and join in the    
 Steiner point which has the property of minimizing the overall length of the   
 three strings. This picture is also in agreement with what one would naively   
 find using standard strong coupling expansion (notice however that due to the   
 roughening transtion this is only a qualitative indication, and cannot be   
 advocated as a``proof'' of the {\bf Y} law). An interesting consequence of this   
 scenario is that  one can use an effective string approach to model the   
 behaviour of these flux tubes and hence a ``L\"uscher like'' correction to the   
 potential should be expected. This correction was evaluated in~\cite{fj03} and   
 succesfully compared with simulations of the 3d ${\mathbb{Z}}_3$ gauge model   
 in~\cite{fj05}.

\item   
 At shorter distances a smooth crossover toward the so called {\bf $\Delta$}   
 law is observed. According to the {\bf $\Delta$} law the three-quark    
 potential is well approximated by the sum of the three two-quark interactions. 
 More precisely the {\bf $\Delta$} law assumes that the three-quark correlator  
 (let us call it $G_3(x_1,x_2,x_3)$ where $x_j$ denotes the position of the $j^{th}$ quark) 
 is related to the quark-antiquark correlator $G_2(x_i,x_j)$ as follows: 
 \eq 
 G_3(x_1,x_2,x_3)\sim \sqrt{G_2(x_1,x_2)~G_2(x_2,x_3)~G_2(x_1,x_3)} 
 \label{delta} 
 \en    
 thus leading to  a potential which increases linearly with the sum of the three   
 interquark distances. The scale where the transition between these two   
 behaviours seems to occur, according to the most recent simulation is around   
 0.8 fm.   
\end{itemize}   
   
To improve our understanding of the baryonic states    
it would be important now to have some quantitative insight in the above    
described picture as well as to have some theoretical argument to explain    
why instead of having a single shape stable for all the interquark distances  a   
$\Delta \to Y$ crossover occurs. Moreover, since the crossover region happens to   
occur exactly in the range of distances which is interesting from a   
phenomenological point of view, it would be important to have some kind of   
theoretical description of this crossover with which to compare the numerical data.   
   
In this respect the present    
study of the three-point function in the 2d ${\mathbb{Z}}_3$ Potts   
model is a perfect laboratory to address this problem. Besides the obvious   
similarity  of the two settings it is also possible to find a direct relation,   
since by dimensional reduction the behaviour at high temperature of the three-quark  
correlator for a $SU(3)$ or a ${\mathbb{Z}}_3$ gauge model in (2+1)   
dimensions is mapped into the behvaiour of the 2d ${\mathbb{Z}}_3$ Potts three-point   
function (in analogy to what happens for the quark-antiquark potential which is   
mappend onto the $\langle\sigma \bar\sigma\rangle$ correlator). 
 
This mapping becomes exact in the vicinity of the deconfinement transition thanks to the fact that both the 
deconfinement transition in the  SU(3) 
LGT in (2+1) dimensions and the magnetization transition of the three-state Potts model in two dimensions are of 
second order. Then, according to the Svetitsky--Yaffe conjecture~\cite{sy82}, the two critical points must 
belong to the same universality class and we can use the three-state Potts model as an effective theory for the 
SU(3) LGT. In this effective description the Polyakov loops of the LGT are mapped onto the spins of the Potts 
model, the confining phase of the LGT into the high temperature phase of the spin model while the combination 
$\sigma/T\equiv \sigma N_t$ (where $T$ is the finite temperature of the LGT model which is equivalent to the 
inverse of $N_t$: the lattice size in the timelike, compactified, direction) is mapped into the mass scale of the 
spin model (i.e.\ the inverse of the correlation length) and sets the scale of the deviations from the critical 
behaviour. It is exactly this scale which separates the {\bf $\Delta$} law behaviour from the {\bf Y} one. 
 
At the critical point (i.e.\ when the correlation length goes to infinity) the {\bf $\Delta$} law is exact. 
In fact looking at eq.(\ref{2pt}) and (\ref{3pt})  
for the conformal two- and three-point correlators we see that the 
relation (\ref{delta}) is fulfilled exactly. On the other side, when the distances among the spins in the 
correlator are much larger than the correlation length, simple strong coupling arguments suggest that the 
dominating configuration in the partition function must be the one which minimizes the distances among the spins 
and the {\bf Y} law appears.  
In field theory this behaviour for large separations among the spins is a direct consequence of the particle fusion  
process (\ref{proc}). Our analysis allows to follow in a rigorous way the crossover between the two 
limiting behaviours. 
 
An interesting and non trivial application    
of our results is that they can give some insight on the high-temperature regime of the string   
corrections (the opposite of the one studied in~\cite{fj05}). This regime is reached when the interquark 
distances 
are much larger than the size of the lattice in the time direction and thus coincides (following the Svetitsky 
Yaffe mapping discussed above) with the large distance limit of the three-point correlator in the Potts model. 
A remarkable and not trivial feature of our result in this limit is that (when the Steiner point lies inside the 
triangle formed by the three spins) the dominating exponential behaviour is 
dressed by a pre-exponential factor $(r_Y)^{-1/2}$ which  
is encoded in the $K_0$ modified Bessel function which appears in 
eq.(\ref{g3symm}). This is exactly the same behaviour of the two-point function in this limit and, as  
for the quark-antiquark case~\cite{lw04,bc05}, it indicates that  
in this limit the effective string corrections give a contribution 
proportional to $\log(r_Y)$ (analogous of the $\log(r)$ term, with $r$ the interquark distance  
in the quark-antiquark case~\cite{lw04,bc05,chp03}). This is a rather non trivial result, which severely  
constraints the possible effective 
string pictures for the three quark potential  
and was indeed observed, in the free effective string limit in the 
case of the ${\mathbb{Z}}_3$ lattice gauge theory in three dimensions~\cite{oj-unp}. 
 
\vspace{2cm}  
{\bf Acknowledgements:} The work of P.G. is supported by the European Commission RTN Network EUCLID (contract  
HPRN-CT-2002-00325); M.C. and G.D. are also partially supported by this contract. O.J.~is supported by funds provided by the U.S.~Department of Energy (D.O.E.) under cooperative research agreement DE-FC02-94ER40818. 
 
\newpage

\section*{Appendix {\bf A}}   
\begin{figure}   
\begin{center}   
\includegraphics[width=14cm]{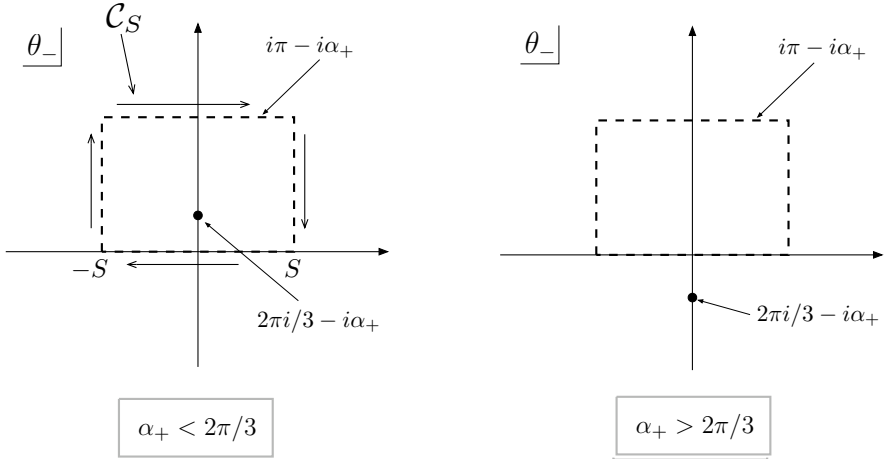}   
\caption{Contour of integration in the $\theta_-$ complex plane.}   
\label{f2}   
\end{center}   
\end{figure}

In order to show explicitly how to obtain equation (\ref{G11}), let us start from (\ref{init})   
and perform the change of variables $\beta \to \beta + i \alpha - i \pi/2$, $\theta \to \theta - i \gamma + i \pi/2$, so  
that we obtain 
\bea   
(F_{\bar A}^\sigma)^2 \int_{-\infty-i \alpha + i \pi/2}^{\infty-i \alpha + i \pi/2} \frac{\textrm{d} \beta}{2 \pi}  \int_{-\infty+i \gamma - i \pi/2}^{\infty+i \gamma - i \pi/2} \frac{\textrm{d} \theta}{2 \pi}    
F_{ A  A}^{ \sigma} (\beta - \theta + i (\alpha + \gamma)) \exp \left[ - m \left( R_{12} \cosh \beta +    
R_{23} \cosh \theta  \right) \right].  
\nonumber   
\eea   
Another change of variables, $\theta_\pm=\beta \pm \theta$,  gives   
\bea   
\frac{(F_{\bar A}^\sigma)^2}{2} \int_{-\infty-i \alpha_-}^{\infty-i \alpha_- } \frac{\textrm{d} \theta_+}{2 \pi} && \hspace{-0.5cm} \int_{-\infty-i \alpha_+ + i \pi}^{\infty-i \alpha_+ + i \pi} \frac{\textrm{d} \theta_-}{2 \pi} \times \nonumber \\   
& &  \hspace{-0.5cm} \times F_{A  A}^{ \sigma} (\theta_- + i \alpha_+ )  
\exp \left[ - m \left( R_{12} \cosh \frac{\theta_+ +\theta_-}{2} +    
R_{23} \cosh  \frac{\theta_+ -\theta_-}{2}  \right) \right] 
\nonumber   
\eea     
where $\alpha_\pm=\alpha \pm \gamma$. Now, we can integrate the function    
\bea   
f(\theta_-) =   
F_{A  A}^{ \sigma} (\theta_- + i \alpha_+ ) \exp \left[ - m \left( R_{12} \cosh \frac{\theta_+ +\theta_-}{2} +    
R_{23} \cosh  \frac{\theta_+ -\theta_-}{2}  \right) \right]   
\eea   
on the complex $\theta_-$-plane along the contour ${\mathcal C}_S$   
depicted in figure \ref{f2}   
\bea   
\oint_{{\mathcal C}_S} f(\theta_-) \, \textrm{d} \theta_- = U_S + D_S + R_S + L_S = - 2 \pi i \, H \left( \frac{2 \pi}{3}-\alpha_+ \right)   
\, \textrm{Res}_{\theta_-=\frac{2 \pi i}{3}-i\alpha_+}   
f(\theta_-)    
\eea   
where $H(z)$ is the usual step-function and   
\bea   
R_S  =  \int_{-S -i \alpha_+ + i \pi}^{S -i \alpha_+ + i \pi} \textrm{d} \theta_-   
\; f(\theta_-), &  \ \ \ \ \ \  &   
L_S =  \int_{-S }^{S} \textrm{d} \theta_-   
\; f(\theta_-) \nonumber \\    
U_S  =  \int_{-S}^{-S -i \alpha_+ + i \pi} \textrm{d} \theta_-   
\; f(\theta_-), &  \ \ \ \ \ \  &   
D_S =  \int_{S -i \alpha_+ + i \pi}^{S} \textrm{d} \theta_-   
\; f(\theta_-)\,. \nonumber     
\eea   
The contribution of the pole is computed using the residue on the bound state   
\bea   
-i \; \textrm{Res}_{\theta_-=\frac{2 \pi i}{3}-i\alpha_+} F_{A A}^{\sigma} (\theta_- + i \alpha_+) = \Gamma^{\bar A}_{AA} F_{\bar A}^\sigma.   
\nonumber   
\eea   
Then, taking the limit $S \to \infty$, both $U_S$ and $D_S$ vanish,    
and we are left with the following expression   
\bea   
&& \hspace{-0.5cm} (F_{\bar A}^\sigma)^2\int_{-\infty}^{\infty} \frac{\textrm{d} \beta}{2 \pi}  \int_{-\infty}^{\infty} \frac{\textrm{d} \theta}{2 \pi}    
\bigg{[} F_{ A  A}^{ \sigma} (\beta - \theta + i (\alpha + \gamma)) \exp \left[ - m \left( R_{12} \cosh \beta +    
R_{23} \cosh \theta  \right) \right] \bigg{]} + \nonumber \\   
&& + H \left( \frac{2 \pi}{3}-\alpha_+ \right) \; \frac{(F_{\bar A}^\sigma)^3}{2} \; \Gamma^{\bar A}_{AA} \int_{-\infty}^{\infty } \frac{\textrm{d} \theta_+}{2 \pi} \times \nonumber \\   
&&\hspace{0.2cm} \times\exp \left[ - m \left( R_{12} \cosh \left( \frac{\theta_+}{2} + \frac{i \pi}{3} -i \alpha  \right) +    
R_{23} \cosh   \left( \frac{\theta_+}{2} - \frac{i \pi}{3} +i \gamma  \right)             \right) \right] \,.  
\nonumber   
\eea    
The desired result can be obtained in the following way. First, the expansion of the argument of the exponential function gives   
\bea   
&& {\mathcal R} = R_{12} \cosh \left( \frac{\theta_+}{2} + \frac{i \pi}{3} -i \alpha  \right) +    
R_{23} \cosh   \left( \frac{\theta_+}{2} - \frac{i \pi}{3} +i \gamma  \right)   = \nonumber \\   
&& = \left[ R_{12} \left( \cos \frac{\pi}{3} \cos \alpha +  \sin \frac{\pi}{3} \sin \alpha \right) +  R_{23} \left(\cos \frac{\pi}{3} \cos \gamma +  \sin \frac{\pi}{3} \sin \gamma \right) \right]   
\cosh  \frac{\theta_+}{2}  + \nonumber \\   
&& +i \left[ R_{12} \left(\sin \frac{\pi}{3} \cos \alpha - \cos \frac{\pi}{3} \sin \alpha \right) -  R_{23} \left(\sin \frac{\pi}{3} \cos \gamma -  \cos \frac{\pi}{3} \sin \gamma \right) \right]   
\sinh  \frac{\theta_+}{2}  ,   
\nonumber   
\eea   
then taking into account the following geometric identities (figure \ref{f1})   
\bea   
&& r_1=R_{12} \frac{\sin \alpha}{\sin \frac{\pi}{3}}, \ \ \ \ r_3=R_{23} \frac{\sin \gamma}{\sin \frac{\pi}{3}}, \nonumber \\   
&& r_2= R_{12} \cos \alpha - r_1 \cos \frac{\pi}{3} = R_{23} \cos \gamma - r_3 \cos \frac{\pi}{3} \nonumber   
\eea   
we finally have    
\bea   
 {\mathcal R} \ = \ ( r_1+r_2+r_3 ) \cosh \frac{\theta_+}{2}  = r_Y \cosh \frac{\theta_+}{2}     
 \nonumber   
\eea   
and the term proportional to $\sinh  \frac{\theta_+}{2}  $ vanishes.   
Hence, by means of these simplifications we are able to obtain formula (\ref{G11}).    
Furthermore, with the same procedure it is possible to compute both eq.(\ref{G12})   
and eq.(\ref{G21}).   
 
\section*{Appendix {\bf B}}   
 
A non-trivial check about the correlation functions written in section \ref{sect3.1} (and as a consequence, about the form factors) is given by the computation of universal ratios. In particular, we are interested in those ratios which involve the  
amplitudes of the high-temperature ($\Gamma_+ $) and ``longitudinal'' ($\Gamma_-$) and ``transverse'' ($\Gamma_{\textrm{\tiny T}}$)  
low-temperature magnetic susceptibilities\footnote{It is worth recalling that the scaling behaviour of the susceptibilities is  
given by   
\bea   
\chi_+ \sim \Gamma _+ \; \tau^{-\gamma}, \ \ \ \  \chi_- \sim \Gamma _- \; |\tau|^{-\gamma},  
\ \ \ \  \chi_{\textrm{\tiny T}} \sim \Gamma _{\textrm{\tiny T}} \; |\tau|^{-\gamma};   
\eea    
where $\gamma=2 \; \frac{1-X_\sigma}{2-X_\epsilon}=\frac{13}{9}$.}.  
These ratios have been calculated in the continuum by  
form factors techniques in the kink basis of the low-temperature phase of the model with the result \cite{Delfino:1997ag,DBC} 
\bea   
\frac{\Gamma_+}{\Gamma_-}\simeq 13.85, \ \ \ \ \ \frac{\Gamma_{\textrm{\tiny T}}}{\Gamma_-}\simeq 0.327.   
\label{ratios} 
\eea   
These values have been confirmed by lattice computations in \cite{Shchur} and \cite{Enting:2003mx}. The most accurate  
lattice estimates come from the latter paper and read $13.83(9)$ and $0.325(2)$, respectively.   
    
Once duality is used to relate correlators in the two phases, the QFT resuls cannot depend on the regime used to compute  
the form factors. Hence, the results (\ref{ratios}) must be reproduced in terms of the high-temperature form factors of the order  
and disorder operators we have used in this paper. Let us recall the definitions of the susceptibilities in terms of the  
correlators:   
\bea   
\chi_+ & = & \int \textrm{d}^2 x \; \langle   \sigma_\alpha (x) \sigma_\alpha (0)  \rangle_{T>T_c} \nonumber \\   
\chi_- & = & \int \textrm{d}^2 x  \; \langle  0_\beta |  \sigma_\beta (x) \sigma_\beta (0) |0_\beta \rangle_{T<T_c}  
\nonumber  \\   
\chi_{\textrm{\tiny T}} & = & \int \textrm{d}^2 x  \; \langle 0_\beta |  \sigma_\alpha (x) \sigma_\alpha (0) |0_\beta  
\rangle_{T<T_c}\,, \ \ \ \ \ \alpha \neq \beta\,.  
\label{susc} 
\eea   
Here  
\begin{equation} 
\sigma_\alpha(x) = \delta_{s(x),\alpha} - \frac13, \ \ \ \ \ \ s(x)=0,1,2   
\end{equation} 
and $|0_\beta\rangle$, $\beta=0,1,2$ are the vacua of the model in the low-temperature phase. The spin variables $\sigma$ and  
$\bar\sigma=\sigma^*$ we used in this paper read 
\bea   
\sigma(x) = e^{s(x) \frac{2 \pi i}{3}}, \ \ \ \ \ \ s(x)=0,1,2   
\eea   
so that 
\bea   
 \sigma_\alpha(x) = \frac13 \left[ \sigma(x) e^{-\alpha \frac{2 \pi i}{3}} + \bar \sigma(x) e^{\alpha \frac{2 \pi i}{3}} \right].   
\eea   
 
The correlators entering (\ref{susc}) can then be expressed in terms of those computed in this paper as 
\bea   
&& \langle \sigma_\alpha (x) \sigma_\alpha (0) \rangle_{T>T_c}=\frac29\; \langle\sigma(x)\bar\sigma(0)\rangle_{T>T_c}  
\label{htcorr}  \\ 
&& \langle 0_\beta|\sigma_\alpha(x)\sigma_\alpha(0)|0_\beta\rangle_{T<T_c}=\frac29\; \langle\mu(x)\bar\mu(0)+\cos\frac{2\pi(\alpha- 
\beta)}{3} \; \mu(x)\mu(0)\rangle_{T>T_c} \,. 
\label{ltcorrs}   
\eea   
In the last equation duality has been used to trade $\sigma$ correlators at low-temperature for $\mu$ correlators at  
high-temperature. We can now use the form factors (\ref{ff}) to compute the susceptibilities up to two-particles   
\bea   
\chi_+ & \simeq & {\mathcal N}\; \frac{\langle \mu \rangle^2}{m^2} \;   0.44674 \nonumber \\   
\chi_- & \simeq & \frac{\langle \mu \rangle^2}{m^2}  \;  0.0312375 \nonumber \\   
\chi_{\textrm{\tiny T}} & \simeq &\frac{\langle \mu \rangle^2}{m^2}   \; 0.0102246 
\eea    
with ${\mathcal N}$ given in (\ref{eq:calN}). Taking the ratios reproduces the results (\ref{ratios}), as expected.

\end{document}